\documentclass[prb,twocolumn,eqsecnum,amsmath,amssymb,showpacs,floatfix,superscriptaddress]{revtex4}

\usepackage{bm}
\usepackage[dvips]{graphicx,color}
\usepackage[mathscr]{eucal}
\usepackage{hyperref}

\allowdisplaybreaks

\newcommand{\be}{\begin{equation}}
\newcommand{\ee}{\end{equation}}
\newcommand{\up}{\uparrow}
\newcommand{\down}{\downarrow}

\begin{document}

\title{Orbital currents and charge density waves in a generalized Hubbard ladder}

\author{J. O. Fj{\ae}restad}
\affiliation{Department of Physics and Astronomy, University of
California, Los Angeles, California 90095, USA}
\affiliation{Department of Physics, University of Queensland,
Brisbane, Qld 4072, Australia}
\author{J. B. Marston}
\affiliation{Department of Physics, Brown University, Providence,
Rhode Island 02912, USA}
\author{U. Schollw\"{o}ck}
\affiliation{Institute for Theoretical Physics C, RWTH Aachen,
D-52056 Aachen, Germany}

\date{\today}

\begin{abstract}
We study a generalized Hubbard model on the two-leg ladder at zero
temperature, focusing on a parameter region with staggered flux
(SF)/d-density wave (DDW) order. To guide our numerical
calculations, we first investigate the location of a SF/DDW phase in
the phase diagram of the half-filled weakly interacting ladder using
a perturbative renormalization group (RG) and bosonization approach.
For hole doping $\delta$ away from half-filling, finite-size
density-matrix renormalization-group (DMRG) calculations are used to
study ladders with up to 200 rungs for intermediate-strength
interactions. In the doped SF/DDW phase, the staggered rung current
and the rung electron density both show periodic spatial
oscillations, with characteristic wavelengths $2/\delta$ and
$1/\delta$, respectively, corresponding to ordering wavevectors
$2k_F$ and $4k_F$ for the currents and densities, where
$2k_F=\pi(1-\delta)$. The density minima are located at the
anti-phase domain walls of the staggered current. For sufficiently
large dopings, SF/DDW order is suppressed. The rung density
modulation also exists in neighboring phases where currents decay
exponentially. We show that most of the DMRG results can be
qualitatively understood from weak-coupling RG/bosonization
arguments. However, while these arguments seem to suggest a
crossover from non-decaying correlations to power-law decay at a
length scale of order $1/\delta$, the DMRG results are consistent
with a true long-range order scenario for the currents and
densities.
\end{abstract}


\maketitle


\section{Introduction}

The possibility of finding new phases of matter is a major
motivation behind the study of materials of strongly correlated
electrons. A phase with staggered orbital currents which was first
considered theoretically in 1968\cite{halric68} and then
rediscovered two decades later
\cite{affmar88,maraff89,nervac89,sch89,hsumaraff91} has lately been
the subject of a revival of interest, mainly due to recent
proposals\cite{chalaumornay01,lee03} that the pseudogap
region\cite{timsta99} in the phase diagram of the cuprate
high-temperature superconductors may be characterized by this kind
of order, either long-ranged (i.e., a true broken-symmetry
state),\cite{chalaumornay01} or fluctuating.\cite{lee03} This
orbital current phase is known variously as the orbital
antiferromagnet, staggered flux (SF) or d-density wave\cite{DWterm}
(DDW) phase; in this paper we will refer to it as the SF/DDW phase.
The long-range ordered version of this phase breaks the rotational
and translational symmetries of the underlying Hamiltonian, in
addition to time reversal symmetry. The fundamental experimental
signature of the long-range-ordered SF/DDW scenario with ordering
wavevector $(\pi,\pi)$ is an elastic Bragg peak at that wavevector
in neutron scattering.\cite{hsumaraff91,chakeenay0102} The results
of some recent neutron scattering experiments on underdoped
YBa$_2$Cu$_3$O$_{6+x}$ have been argued to be consistent with this
scenario.\cite{ mooetal0102,chakeenay0102,mooetal04} A different
circulating-current broken-symmetry state with a current pattern
that does not break translational symmetry has also been proposed
for the pseudogap phase in the cuprates.\cite{varsim}

It is important to acquire an understanding of what kinds of
microscopic models can give rise to a SF/DDW ground state. For
models of interacting fermions in two spatial dimensions this issue
has been addressed by many
authors\cite{affmar88,maraff89,nervac89,sch89,hsumaraff91,capzey99,ivaleewen00,nay00,
leu00,staphi01,naypiv02,binbaedou02,kamkat03,chukeekim03,norole04,ass04}
using a variety of methods. While the two-dimensional case is the
most relevant one for the cuprates, the behavior of strongly
correlated electrons in two dimensions is still a subject that is
marred by great controversy. In a few special model cases the
fermion sign problem is absent and thus quantum Monte Carlo
simulations can be done reliably, leading among other things to
interesting findings with regard to SF/DDW order\cite{ass04} and
also more exotic current-carrying states.\cite{capwuzha04} However,
in the vast majority of cases, and certainly for the models and
parameter values that are expected to be most relevant for real
materials, the available analytical and numerical methods are only
approximate, and their reliability is difficult to gauge.

In contrast, the fact that many powerful methods exist for one
spatial dimension has enabled much solid knowledge to be established
about strongly correlated one-dimensional systems, and it is hoped
that some of this may also be relevant for the behavior of
correlated electrons in two dimensions. In this regard, two-leg
Hubbard and $t$-$J$ ladders have attracted much interest, as it has
been found that these models have a spin gap and that upon doping
away from half-filling the dominant correlations are $d$-wave-like
superconducting correlations,\cite{dagric} two features which are
reminiscent of the pseudogap and the $d_{x^2-y^2}$ symmetry of the
superconducting state in the cuprates, respectively. Ladder systems
are also interesting in their own right, not least due to
experimental realizations of such materials. The most well-known
example, Sr$_{14-x}$Ca$_x$Cu$_{24}$O$_{41}$, contains two-leg ladder
substructures, and has been found to have a spin gap, and to become
superconducting upon doping at high pressure.\cite{uehetal96,dagric}

The possibility of having SF/DDW order in two-leg ladders has also
received much attention
recently.\cite{ner91,nerlutkus93,sch96,origia97,scawhiaff01,tsupoicap01,fjamar02,marfjasud02,
wuliufra03,tsufur02,schetal03,momhik03,momhik04} The two-leg ladder is the
simplest geometry that can support a SF/DDW phase, with currents
flowing around the elementary square plaquettes. In this paper we
present an extensive study of SF/DDW order in generalized Hubbard
ladders, focusing mainly on the case of finite hole doping away
from half-filling. Some of the results that will be discussed have been
briefly presented in Ref. \onlinecite{schetal03}.

We first use bosonization and a perturbative renormalization-group (RG)
approach to identify a parameter region with long-range SF/DDW order
in the weakly interacting half-filled
ladder.\cite{fjamar02,tsufur02,wuliufra03} Finite-system
density-matrix renormalization-group (DMRG) calculations are then
used to study the SF/DDW phase in doped ladders, for
intermediate-strength interactions and ladder sizes up to 200 rungs.
For the (rational) hole dopings considered, currents are found to be
large (of order the nearest-neighbor hopping amplitude $t$) and show
no evidence of decay. As the doping is increased the currents in the
SF/DDW phase decrease in magnitude, and for sufficiently strong
doping SF/DDW order becomes completely suppressed. The SF/DDW phase
has a spin gap, but $d$-wave-like superconducting correlations decay
exponentially.

In the half-filled SF/DDW phase, the sign of the rung current
changes from one rung to the next, while the magnitude stays the
same, corresponding to an ordering wavevector $2k_F=\pi$ in the
direction parallel to the legs. In contrast, the rung currents found
in the doped SF/DDW phase vary both in sign and magnitude,
corresponding to an ordering wavevector $2k_F=\pi(1-\delta)\neq
\pi$, where $\delta$ is the hole doping away from half-filling. Such
phases are often referred to as incommensurate. An incommensurate
SF/DDW phase has been suggested as a candidate for the hidden order
in the heavy fermion compound URu$_2$Si$_2$.\cite{chacolmydtri02}
Incommensurate SF/DDW phases have also been studied in various
two-dimensional models.\cite{capzey99,keekim02,chukeekim03} [A note
on terminology: In the remainder of this paper only ordering
patterns with an infinite unit cell will be characterized as
`incommensurate;' orderings with a finite unit cell (even when it is
very large) will be referred to as `commensurate.' The former
(latter) case includes orderings with wavevector
$2k_F=\pi(1-\delta)$ with $\delta$ an irrational (rational) number.]

Charge density wave order in ladder systems is also a topic that has
attracted much interest recently, both
theoretically\cite{vojhubnoa01,whiaffsca02,tsufur02,oricit03,wuliufra03,schwer04,tsusuz04}
and
experimentally.\cite{goretal02,bluetal02,vuletal03,hesetal04,vuletal04,abbetal04}
Interestingly, we find that a charge density modulation coexists
with the orbital currents in the doped SF/DDW phase. Like the
currents, this density modulation is not found to decay. The
staggered rung current and the rung electron density both show a
periodic spatial variation, with characteristic wavelengths
$2/\delta$ and $1/\delta$, respectively. The density modulation
corresponds to an ordering wavevector $4k_F$, with two doped holes
(one per leg) per wavelength. Furthermore, the minima of the
electron density are located at the zeros of the staggered rung
current. In other words, the rungs at which the doped holes
predominantly sit are antiphase domain walls for the current
pattern. This property, and the related factor of two ratio between
the periods of the staggered rung current and the rung electron
density, are very reminiscent of the properties of stripe phases
with coexisting spin and charge order in doped
antiferromagnets,\cite{caremekivorg02} in which the role of the
current in the SF/DDW phase is played by the spin density. The rung
density modulation is found to persist in the neighboring phases of
the SF/DDW phase in the phase diagram in which current correlations
decay exponentially. We note that coexistence of current and charge
order for an SF/DDW phase in two dimensions with ordering wavevector
$\bm{q}\neq(\pi,\pi)$ for the currents has previously been discussed
within the framework of a phenomenological mean-field
approach.\cite{keekim02}

Is the current and charge density order long-ranged? The DMRG
calculations do seem to support a commensurate true long-range order
scenario. Let us first note that, to the best of our knowledge, such
a scenario does not violate any exact theorems that are directly
applicable to the lattice model studied here; in particular, the
Mermin-Wagner theorem is respected. Of course, these nonviolation
arguments do not say whether or not true long-range order actually
will occur. A weak-coupling RG/bosonization analysis of the
continuum limit of the lattice model predicts that (1) true
long-range order requires Umklapp interactions to be relevant in the
RG sense, (2) at half-filling, Umklapp interactions are relevant,
resulting in true long-range
order,\cite{fjamar02,tsufur02,wuliufra03} (3) for dopings away from
half-filling Umklapp interactions are irrelevant, and as a result
correlation functions decay as power laws (for distances much larger
than $1/\delta$\cite{wuliufra03}), i.e., only quasi long-range order
is predicted.\cite{fjamar02,wuliufra03} DMRG calculations find true
long-range order at half-filling,\cite{marfjasud02} thus agreeing
with the RG/bosonization analysis in that case. But as already
noted, DMRG calculations seem to support a true long-range order
scenario also for the doped ladder. We have not been able to resolve
this apparent discrepancy between the predictions of the DMRG and
RG/bosonization methods, and can only speculate about its origin.
Both methods have their limitations and weaknesses. One particular
question mark associated with the weak-coupling RG/bosonization
method is whether its predictions regarding the range of the order
may be unreliable for the ladders studied here with DMRG, for which
the interactions are not weak. But one certainly cannot exclude the
possibility that there may be a different and perhaps quite subtle
reason for the discrepancy. We refer the reader to the concluding
Sec. \ref{conclusion} for an attempt at a fuller discussion of these
issues.

The paper is organized as follows. The generalized Hubbard model on
the two-leg ladder is introduced in Sec. \ref{model}. In Sec.
\ref{weakint} we use a perturbative RG approach
and bosonization to study the phase diagram of the model for weak
interactions at half-filling. In Sec. \ref{dmrg} we present an
extensive DMRG study of the model for nonzero doping and
intermediate-strength interactions. In Sec. \ref{bosdoped} the doped
ladder is discussed using weak-coupling RG and bosonization
arguments. Sec. \ref{conclusion} contains a summary and concluding
discussion. Some details of the RG/bosonization calculations are
discussed in three appendices. Throughout the paper the main focus
is on the SF/DDW phase, but properties of neighboring phases in the
phase diagram are also discussed.

\section{Model}
\label{model}

The model we will study is an ``extended''or ``generalized'' Hubbard
model with various nearest-neighbor interactions in addition to the
on-site Hubbard term. For the purpose of finding SF/DDW order, it is
necessary to allow for these additional interactions because the
ground states of the pure Hubbard ladder and the related $t$-$J$
ladder have been found to only have short-ranged SF/DDW
correlations.\cite{scawhiaff01,marfjasud02} Another, more general,
motivation for studying generalized Hubbard models is that, although
much attention has been paid over the years to the two-dimensional
Hubbard and $t$-$J$ models as purportedly ``minimal'' effective
models for describing the physics of the two-dimensional CuO$_2$
planes in the cuprate superconductors, there is an increasing amount
of studies that suggest that these models do not support high-temperature
superconductivity.\cite{beyondhubbard,prykivzac03}

The Hamiltonian of our model is $H=H_0+H_I$, where the kinetic
energy and interaction operators are, respectively,
\begin{eqnarray}
H_0 &=& -t\sum_{xs}\bigg(\sum_{\ell}c^{\dagger}_{\ell,x+1,s}c_{\ell xs}
 + c^{\dagger}_{txs}c_{bxs}\bigg)+\text{H.c.},
\label{KE} \nonumber \\
H_I &=& \sum_{x} \bigg(\sum_{\ell}\big[U n_{\ell,x,\uparrow}n_{\ell,x,\downarrow} +
V_{\parallel}n_{\ell x}n_{\ell,x+1} \nonumber \\ & & \hspace{-1cm} +
\; J_{\parallel}\bm{S}_{\ell x}\cdot\bm{S}_{\ell,x+1}\big]+ V_{\perp}n_{tx}n_{bx}
+ J_{\perp}\bm{S}_{tx}\cdot\bm{S}_{bx}\bigg).
\label{HI-latt}
\end{eqnarray}
The operator $c^{\dagger}_{\ell xs}$ creates an electron on rung
$x=1,\ldots,L$ on the ``top'' or ``bottom'' leg $\ell=t,b$ with spin
$s=\uparrow,\downarrow$, and obeys $\{c_{\ell
xs},c^{\dagger}_{\ell'x's'}\}=\delta_{\ell\ell'}\delta_{xx'}\delta_{ss'}$.
Electrons can hop between nearest-neighbor sites along legs and
rungs with hopping amplitude $t$. Electron density operators are
given by $n_{\ell xs}=c^{\dagger}_{\ell xs}c_{\ell xs}$ and $n_{\ell
x}=\sum_{s}n_{\ell xs}$, and the spin operator is $\bm{S}_{\ell
x}=\frac{1}{2}\sum_{ss'}c^{\dagger}_{\ell xs}\bm{\sigma}_{ss'}
c_{\ell xs'}$, where the $\sigma^i$ are Pauli matrices. The
interactions consist of the Hubbard on-site term with strength $U$,
as well as nearest-neighbor density-density and spin exchange
interactions with strength $V_{\perp}$ and $J_{\perp}$ along the
rungs, and $V_{\parallel}$ and $J_{\parallel}$ along the legs.
As the leg interactions will be seen to favor other phases than
SF/DDW, we will in most of the paper take $V_{\parallel}=J_{\parallel}=0$.

\section{Weakly interacting electrons on the half-filled ladder}
\label{weakint}

In this section we study the case of weakly interacting electrons on
a half-filled ladder.\cite{linbalfis98,fjamar02,wuliufra03,tsufur02}
The fermionic perturbative RG approach of Refs.
\onlinecite{balfis96,linbalfis97,linbalfis98} is employed to obtain
a low-energy effective theory which is then analyzed using
bosonization and semiclassical considerations. Our focus
is on finding a parameter regime for which the model
(\ref{HI-latt}) has a ground state with long-ranged SF/DDW order.
The phase diagrams obtained for the weakly interacting half-filled
ladder serve as helpful guides in the search for SF/DDW order
in the doped ladder for intermediate interaction strengths, using
the DMRG method, as described in Sec. \ref{dmrg}. Furthermore, some
of the results established here are used in the
RG/bosonization analysis of the doped ladder in Sec. \ref{bosdoped}.

\subsection{Continuum description}

To study the weak-interaction problem, we first diagonalize the
kinetic energy. As it is convenient to deal with a translationally
invariant Hamiltonian, we impose periodic boundary conditions in the
direction parallel to the legs.  By introducing even and odd
combinations $c_{\lambda xs}=(c_{txs} +
(-1)^{\lambda}c_{bxs})/\sqrt{2}$ with $\lambda=1,2$, and Fourier
transforming along the leg direction, the kinetic energy is
diagonalized in momentum space, describing an anti-bonding band
($\lambda=1$) and a bonding band ($\lambda=2$), with dispersions
$\varepsilon_{\lambda}(k)=-2t\cos k - (-1)^{\lambda}t$ (we set the
lattice constant equal to 1). The Fermi level of the noninteracting
half-filled system is at zero energy and crosses both bands, thus
giving rise to four Fermi points $\pm k_{F1,2}$ which satisfy
\begin{equation}
k_{F1}+k_{F2}=\pi.
\label{kFsum}
\end{equation}
As interactions are assumed to be weak, we focus on states very
close to the Fermi energy, and linearize the kinetic energy around
the Fermi points. The band operator $c_{\lambda xs}$ can be
expressed as a sum of right-moving (R) and left-moving (L)
components,
\be
c_{\lambda xs}=\sum_{P}
e^{iPk_{F\lambda}x}\psi_{P\lambda s}(x),
\label{RLdecompose}
\ee
where $P=R,L\equiv \pm 1$. The continuum fields $\psi_{P\lambda
s}(x)$ are slowly varying on the scale of the lattice constant. The
linearized kinetic energy can be written $H_0= \int dx{\cal H}_0$,
where
\be
{\cal H}_0 = -i v_F \sum_{P\lambda s} P\;
\psi^{\dagger}_{P\lambda s}\partial_x\psi_{P\lambda s},
\label{KEdensity}
\ee
where the Fermi velocity is $v_F=\sqrt{3}t$.

Interactions will scatter electrons between the Fermi points. Introducing
the ``currents''\cite{balfis96}
\begin{eqnarray}
\lefteqn{
J_{\lambda P}=\sum_{s}\psi^{\dagger}_{P\lambda s}
\psi_{P\lambda s},\;
\bm{J}_{\lambda P}=\frac{1}{2}\sum_{ss'}
\psi^{\dagger}_{P\lambda s}\bm{\sigma}_{ss'}
\psi_{P\lambda s'},} \nonumber \\
 & & L_P=\sum_{s}\psi^{\dagger}_{P1s}\psi_{P2s},\quad
 \bm{L}_P=\frac{1}{2}\sum_{ss'}
\psi^{\dagger}_{P1s}\bm{\sigma}_{ss'}\psi_{P2s'}, \nonumber \\
 & & M_{\lambda P}=-i\psi_{P\lambda\up}\psi_{P\lambda \down},\quad
  N_{Pss'}=\psi_{P1s}\psi_{P2s'},
\end{eqnarray}
the Hamiltonian density that describes the effects of the weak interactions
to leading order can be written $\mathcal{H}_I=\mathcal{H}_I^{(1)}+\mathcal{H}_I^{(2)}$,
where
\begin{eqnarray}
\text{$\mathcal{H}$}^{(1)}_I &=& - g_{1\rho}\sum_{\lambda}J_{\lambda R}J_{\lambda L}
- g_{x\rho}(J_{1R}J_{2L}+J_{2R}J_{1L}) \nonumber \\ &-&
 g_{1\sigma}\sum_{\lambda}\bm{J}_{\lambda R}\cdot\bm{J}_{\lambda L} -
g_{x\sigma} (\bm{J}_{1R}\cdot\bm{J}_{2L}+\bm{J}_{2R}\cdot\bm{J}_{1L})
\nonumber \\
 & & \hspace{-1.2cm} - \; g_{t\rho}(L_R L_L + L_R^{\dagger}L_L^{\dagger})
 - g_{t\sigma}(\bm{L}_R\cdot\bm{L}_L + \bm{L}_R^{\dagger}\cdot
\bm{L}_L^{\dagger}),\\
\label{momconsintden}
\text{$\mathcal{H}$}^{(2)}_I &=& - g_{xu}(M_{1R}M_{2L}^{\dagger}+
M_{2R}M_{1L}^{\dagger}) \nonumber \\ & & \hspace{-1.0cm} - \;\sum_{ss'}(g_{tu1}N_{Rss'}
N_{Lss'}^{\dagger}+ g_{tu2}N_{Rss'}N_{Ls's}^{\dagger}) + \mbox{H.c.}
\label{umklappintden}
\end{eqnarray}
Here $\mathcal{H}_I^{(2)}$ represents (interband) Umklapp interactions.

\subsection{Bosonization}

Bosonization proves very helpful for interpreting the low-energy
effective theory resulting from the perturbative RG flow (see Sec.
\ref{rg}). In the Abelian bosonization
formalism\cite{schcunpie98,delsch98,gognertsv98,sen04,gia03} the
fermionic field operators $\psi_{P\lambda s}$ can be expressed in
terms of dual or conjugate Hermitian bosonic fields $\phi_{\lambda
s}$ and $\theta_{\lambda s}$ as
\begin{equation}
 \psi_{P\lambda s}= \frac{1}{\sqrt{2\pi\alpha}}\eta_{\lambda s}
\exp{[i(P\phi_{\lambda s}+\theta_{\lambda s})]}, \label{bosfieldop}
\end{equation}
where $\alpha$ is a short-distance cutoff. The only equal-time
nonzero commutator between the bosonic fields is (taking the limit
$\alpha/(x-x')\to 0$)
\begin{equation}
\text{}[\phi_{\lambda s}(x),\theta_{\lambda's'}(x')] = i\pi
\delta_{\lambda\lambda'}\delta_{ss'}H(x-x'),
\end{equation}
where $H(x)$ is the Heaviside step function. The long-wavelength
normal-ordered fermionic densities can be expressed in terms of the
bosonic fields as
\begin{equation}
\psi^{\dagger}_{P\lambda s}\psi_{P\lambda s} = \frac{1}{2\pi}\partial_x
(\phi_{\lambda s}+P\theta_{\lambda s}).
\label{longwdens}
\end{equation}
For later use we also define
\begin{equation}
{\cal N}_{P\lambda s}\equiv \int_0^L dx\; \psi^{\dagger}_{P\lambda s}\psi_{P\lambda s}.
\end{equation}
The (Majorana) Klein factors\cite{schcunpie98,sen04} $\eta_{\lambda
s}$ in Eq. (\ref{bosfieldop}) commute with the bosonic fields, and
satisfy $\{\eta_{\lambda s},\eta_{\lambda's'}\}=
2\delta_{\lambda\lambda'}\delta_{ss'}$. The Klein factor conventions
used for the Hamiltonian and the order parameters considered in this
paper (see below) are explained in Appendix \ref{kleinconv}.

Charge $(\rho)$ and spin $(\sigma)$ operators are defined as
$\phi_{\lambda\rho} = (\phi_{\lambda\up}
+\phi_{\lambda\down})/\sqrt{2}$, $\phi_{\lambda\sigma} =
(\phi_{\lambda\up}-\phi_{\lambda\down})/\sqrt{2}$. It is convenient
to make a further change of basis, to
$\phi_{r\nu}=(r\phi_{1\nu}+\phi_{2\nu})/\sqrt{2}$, where $r=\pm$ and
$\nu=\rho,\sigma$. Similar definitions apply to the
$\theta$-operators. In the $(r,\nu)$-basis, the kinetic energy
density reads
\begin{equation}
{\cal H}_0 = \frac{v_F}{2\pi}\sum_{r\nu}\left[(\partial_x
\phi_{r\nu})^2+(\partial_x\theta_{r\nu})^2\right]. \label{H0boson}
\end{equation}
Furthermore, ${\cal H}_I^{(1)}={\cal H}_I^{(1a)}+{\cal H}_I^{(1b)}$,
where
\begin{eqnarray}
{\cal H}_{I}^{(1a)} &=& \frac{1}{2\pi^2}\sum_{r\nu}g_{r\nu}
\left[(\partial_x\phi_{r\nu})^2-(\partial_x
\theta_{r\nu})^2\right],
\label{HI1a} \\
\lefteqn{{\cal H}_I^{(1b)} = \frac{1}{(2\pi\alpha)^2}\Big\{2\cos
2\phi_{+\sigma}\Big[g_{1\sigma}\cos 2\phi_{-\sigma}} \nonumber \\ &+&
 g_{x\sigma}\cos 2\theta_{-\sigma} - g_{t\sigma}\cos 2\theta_{-\rho}\Big]
 - \cos 2\theta_{-\rho} \nonumber \\ & & \hspace{-1.8cm}\Big[
(g_{t\sigma}-4g_{t\rho})\cos 2\phi_{-\sigma} + (g_{t\sigma}+4g_{t\rho})\cos
2\theta_{-\sigma}\Big]\Big\},
\label{HI1b}
\end{eqnarray}
with $g_{r\rho}\equiv -\left(g_{1\rho}+rg_{x\rho}\right)$ and $g_{r\sigma}
\equiv -\frac{1}{4}(g_{1\sigma}+rg_{x\sigma})$. Finally, the Umklapp
interaction density reads
\begin{eqnarray}
{\cal H}_I^{(2)} &=& \frac{4}{(2\pi\alpha)^2} \cos 2\phi_{+\rho}\Bigl[
g_{xu}\cos 2\theta_{-\rho} + g_{tu1}\cos2\phi_{-\sigma} \nonumber \\ &+&
(g_{tu1}+g_{tu2})\cos 2\phi_{+\sigma} + g_{tu2}\cos2\theta_{-\sigma}\Bigr].
\label{HI2}
\end{eqnarray}

\subsection{Perturbative renormalization group approach}
\label{rg}

The ``bare'' values of the nine coupling constants $g_i$ in the
continuum field theory can be expressed in terms of the interaction
parameters in Eq. (\ref{HI-latt}), and equations describing the RG
flow (to one loop) of these coupling constants near the
noninteracting fixed point can be derived\cite{balfis96} (see
Appendix \ref{apprg}). These RG equations are then solved
numerically for given sets of initial conditions. Typically, it is
found that as the length scale is increased, some of the couplings
grow (sometimes after a sign change), while other couplings remain
small.

The couplings which grow to be large (i.e., of order 1, at which
point the numerical integration must be stopped since the
perturbative RG equations are only valid for weak coupling) will
tend to lock some of the bosonic fields in the minima of the cosine
potentials in the Hamiltonian. In fact, in the parameter region
considered in this paper, at half-filling one of the conjugate
fields $\phi_{r\nu}$ and $\theta_{r\nu}$ in each sector $r\nu$
always becomes locked. These locked (and thus also gapped) operator
fields can then be approximated by their c-number expectation values
(integer multiples of $\pi/2$ for the Hamiltonian considered above),
while their conjugate fields will fluctuate wildly. To deduce the
nature of the ground state, it is necessary to bosonize various
order parameters and use the information about the field-locking
pattern to deduce which, if any, of these order parameters are
nonzero, or have the most slowly decaying correlations.

\subsection{Phases, physical observables, and order parameters}
\label{OPs}

\begin{figure}
\centerline{\includegraphics[scale=0.32]{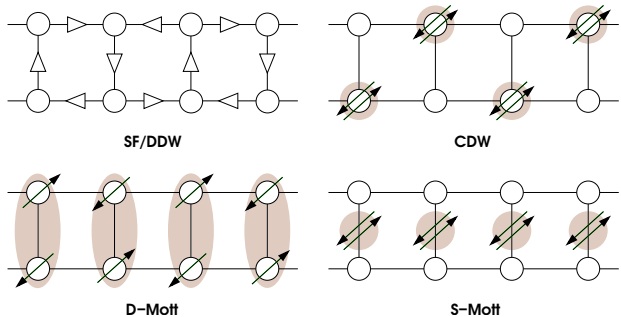}}
\caption{Schematic illustration of phases at half-filling in the
strong-coupling limit. Open (filled) arrows represent currents
(spins). The shading indicates singlet pairs.} \label{picphases}
\end{figure}

As our main interest here is the SF/DDW phase, we will focus on a
parameter region where this kind of order is realized. Three other
phases will also be seen to be present in this parameter region in
the weakly interacting half-filled ladder.  These are the CDW,
D-Mott, and S-Mott phases.\cite{phasenames} The nature of these four
phases, most easily understood in the strong-coupling
limit,\cite{trotsuric96,scazhahan98,linbalfis98,tsufur02} is
illustrated schematically in Fig. \ref{picphases}. These phases are
all completely gapped, i.e., both charge sectors $\pm\rho$ and both
spin sectors $\pm\sigma$ are gapped. Recently it has been shown that
the two-leg ladder also can support four other completely gapped
phases at half-filling,\cite{tsufur02,wuliufra03} but none of these
additional phases appear in the parameter region studied here.

We now discuss in some detail the various phases and the order
parameters/physical observables that characterize them, \textit{at
half-filling}\cite{linbalfis98,fjamar02,tsufur02,wuliufra03} (the
bosonized form of these physical observables in the \textit{doped}
ladder will be discussed in Sec. \ref{physobsdoped}). In the SF/DDW
phase, currents flow around the plaquettes as shown in Fig.
\ref{picphases}. The ordering wavevector of the currents is
$(q_x,q_y)=(\pi,\pi)$, where $q_x$ ($q_y$) is the wavevector in the
direction parallel (perpendicular) to the legs. The order parameter
of this phase is
\begin{eqnarray}
O_{\text{SF/DDW}} &=& \cos\phi_{+\rho}\cos \phi_{+\sigma}\cos\theta_{-\rho}\cos\theta_{-\sigma}
\nonumber \\ &+& \sin\phi_{+\rho}\sin\phi_{+\sigma}\sin\theta_{-\rho}\sin\theta_{-\sigma}.
\label{SF/DDW-OP}
\end{eqnarray}
Denoting the rung current by $j_{\perp}(x)$, and defining the
{\em staggered} rung current as
\begin{equation}
j_s(x)\equiv (-1)^x j_{\perp}(x),
\label{js}
\end{equation}
we have $\langle j_s(x)\rangle\propto \langle
O_{\text{SF/DDW}}(x)\rangle$; note that the expectation values are
independent of $x$.

In the CDW phase the electron density exhibits a checkerboard
pattern. This charge density order is thus characterized by the same
ordering wavevector $(\pi,\pi)$ as the currents in the SF/DDW phase.
The order parameter is
\begin{eqnarray}
O_{\text{CDW}} &=& \cos\phi_{+\rho}\cos \phi_{+\sigma}\sin \theta_{-\rho}\cos\theta_{-\sigma} \nonumber \\
 &-& \sin\phi_{+\rho}\sin\phi_{+\sigma}\cos\theta_{-\rho}\sin\theta_{-\sigma}.
\label{CDW-OP}
\end{eqnarray}
In the CDW phase the expectation value of the deviation $\delta
n_{\ell}(x)$ of the charge density at leg $\ell$, rung $x$ from the
average density is given by $\langle \delta n_{t,b}(x)\rangle
\propto \pm (-1)^{x}\langle O_{\text{CDW}}(x)\rangle$.

Both the SF/DDW and CDW phase spontaneously break a $Z_2$ symmetry,
and thus have a two-fold degenerate ground state with true
long-range order. This is intimately related to the fact that the
order parameters of these phases depend on $\phi_{+\rho}$ which is
locked by the Umklapp interaction, Eq. (\ref{HI2}). In contrast, the
D-Mott and S-Mott phases do not break any symmetry, but are
characterized by a unique ground state with exponentially decaying
$d$-wave-like and $s$-wave-like superconducting (SC) correlations,
respectively. The pairing operators can be taken to be
\begin{eqnarray}
\Delta_{\text{DSC/SSC}}(x) &=& (\psi_{R1\up}\psi_{L1\down}
+\psi_{L1\up}\psi_{R1\down}) \nonumber \\
 &\mp & (\psi_{R2\up}\psi_{L2\down}+\psi_{L2\up}\psi_{R2\down}).
\label{DSC-SSC-PO}
\end{eqnarray}
In $\Delta_{\text{DSC}}$ the components of bands 1 and 2
(corresponding to transverse wavevector $\pi$ and $0$, respectively)
come with opposite signs, while in $\Delta_{\text{SSC}}$ they have
the same sign; this justifies calling these operators $d$- and
$s$-wave-like, respectively. One finds that $\langle
\Delta_a(x)\Delta^{\dagger}_a(0)\rangle\propto \langle
O_a(x)O^{\dagger}_a(0)\rangle$, where $a=$ DSC or SSC, and the order
parameters are
\begin{eqnarray}
O_{\text{DSC}} &=& e^{i\theta_{+\rho}}(\cos\phi_{+\sigma}\cos\theta_{-\rho}\cos\phi_{-\sigma} \nonumber \\
 & & -i\;\sin\phi_{+\sigma}\sin\theta_{-\rho}\sin\phi_{-\sigma}), \label{DSC-OP} \\
O_{\text{SSC}} &=& e^{i\theta_{+\rho}}(\sin\phi_{+\sigma}\cos\theta_{-\rho}\sin\phi_{-\sigma} \nonumber \\
 & & -i\;\cos\phi_{+\sigma}\sin\theta_{-\rho}\cos\phi_{-\sigma}). \label{SSC-OP}
\end{eqnarray}
The exponential decay of the pairing correlations comes from the
$e^{i\theta_{+\rho}}$ factor, as $\theta_{+\rho}$ is conjugate to
the locked field $\phi_{+\rho}$.

The field-locking patterns that realize the SF/DDW and CDW phases
are easily deduced from their respective order parameters. Due to
the absence of $\phi_{+\rho}$ in the DSC and SSC order parameters,
the value of $\langle\phi_{+\rho}\rangle$ in the D-Mott and S-Mott
phases must either be determined from the RG flow, or it can be
deduced from the fact that the SF/DDW and the CDW phase can be
regarded as the Ising-ordered counterparts of the quantum disordered
D-Mott and S-Mott phase, respectively.\cite{tsufur02} Field-locking
patterns for the four phases are given in Table \ref{phasetable}.
\begin{table}
\begin{center}
\begin{tabular}{|l|c|c|c|c|c|}\hline
Phase &  $\langle \phi_{+\rho}\rangle$ & $\langle\phi_{+\sigma}\rangle$ & $\langle\theta_{-\rho}\rangle$ &
$\langle\theta_{-\sigma}\rangle$ & $\langle\phi_{-\sigma}\rangle$
\\\hline
SF/DDW & 0, $\pi$ & 0 & 0 & 0 & $\sim$ \\
CDW & 0, $\pi$ & 0 & $\pi/2$ & 0 & $\sim$ \\
D-Mott  & 0 & 0 & 0 & $\sim$ & 0 \\
S-Mott & 0 & 0 & $\pi/2$ & $\sim$ & 0 \\\hline
\end{tabular}
\caption{Field-locking patterns (only unique up to a global gauge
transformation of the fields) characterizing the ground states of
the four phases encountered at half-filling. The two-fold degeneracy
of the SF/DDW and CDW ground states has been reflected in the two
values given for $\langle \phi_{+\rho}\rangle$. As
$\theta_{-\sigma}$ and $\phi_{-\sigma}$ are conjugate fields, it
follows that if one of them is locked, the other is strongly
fluctuating; this is indicated by `$\sim$'.}
\label{phasetable}
\end{center}
\end{table}

\subsection{Weak-coupling phase diagrams}
\label{phasediagrams}

\begin{figure}[!htb]
\centerline{\includegraphics[scale=0.70,angle=0]{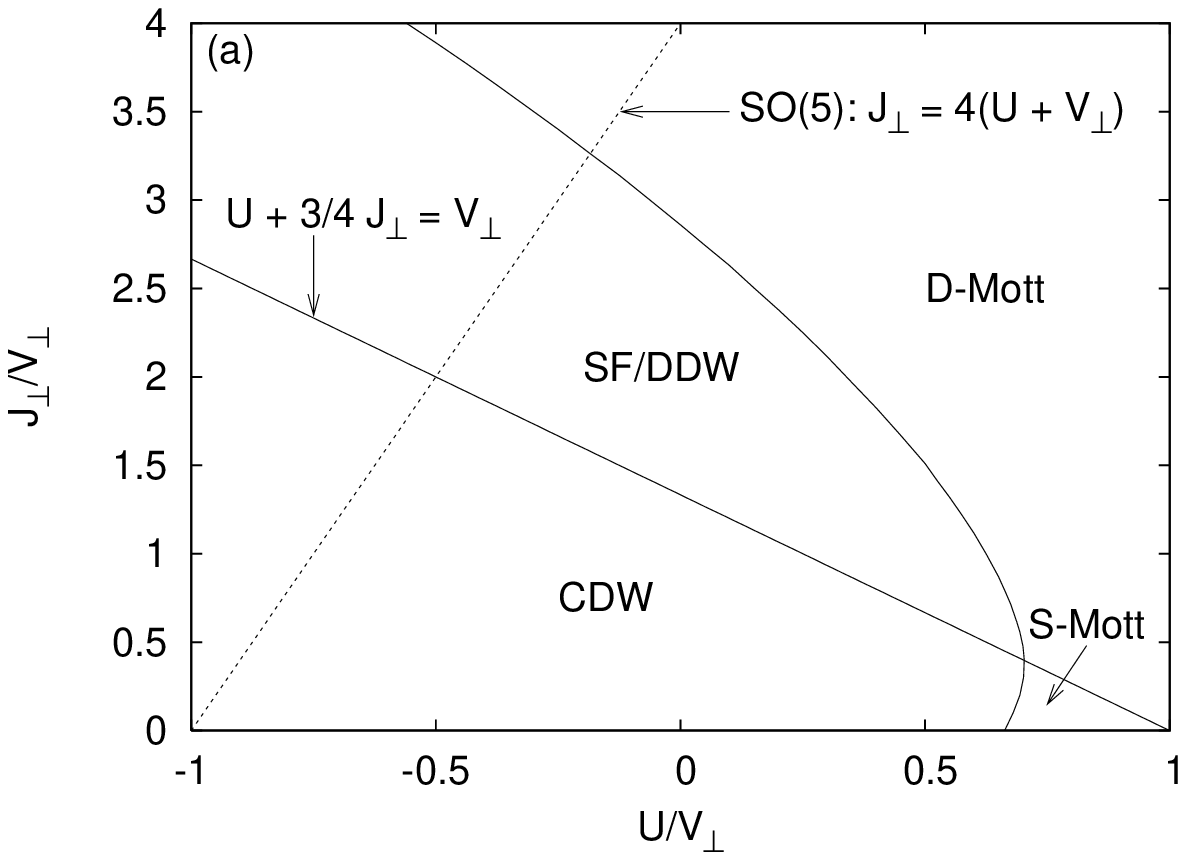}}
\centerline{\includegraphics[scale=0.70,angle=0]{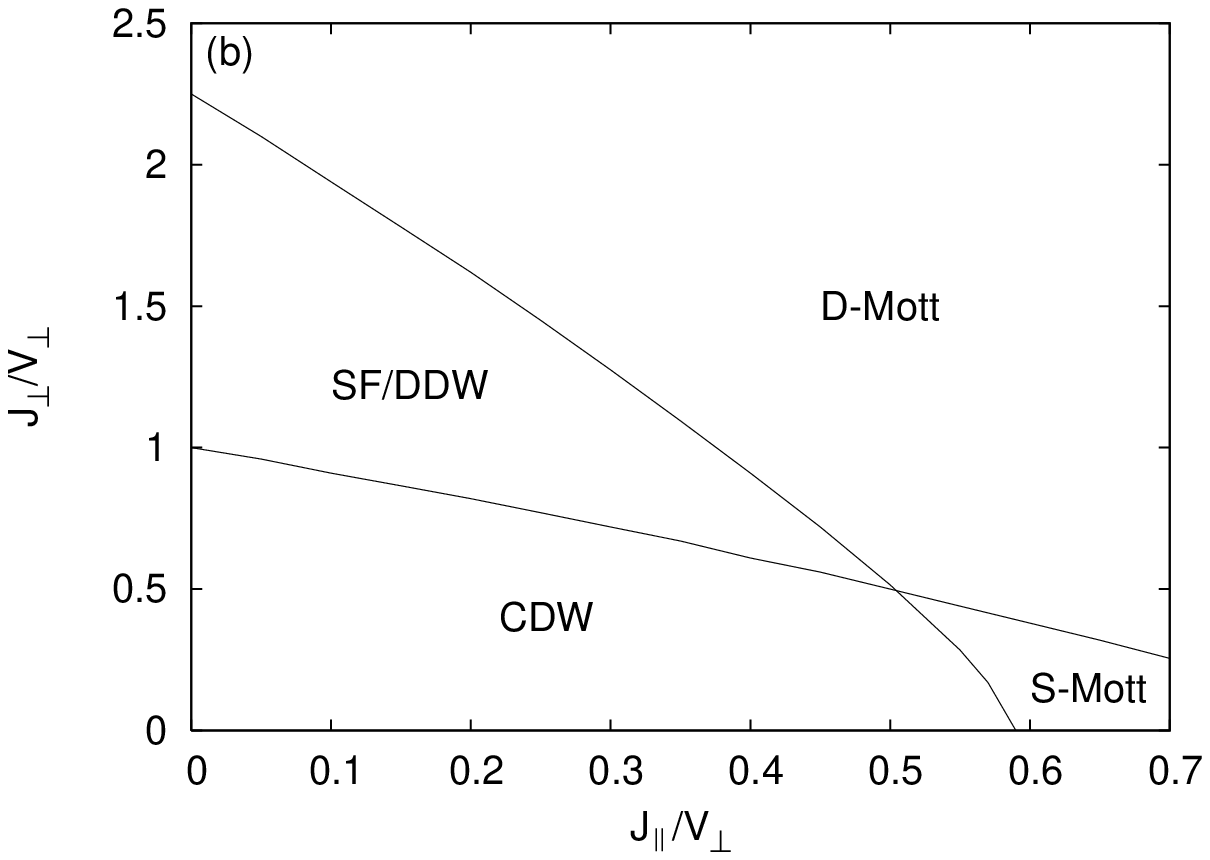}}
\centerline{\includegraphics[scale=0.67,angle=0]{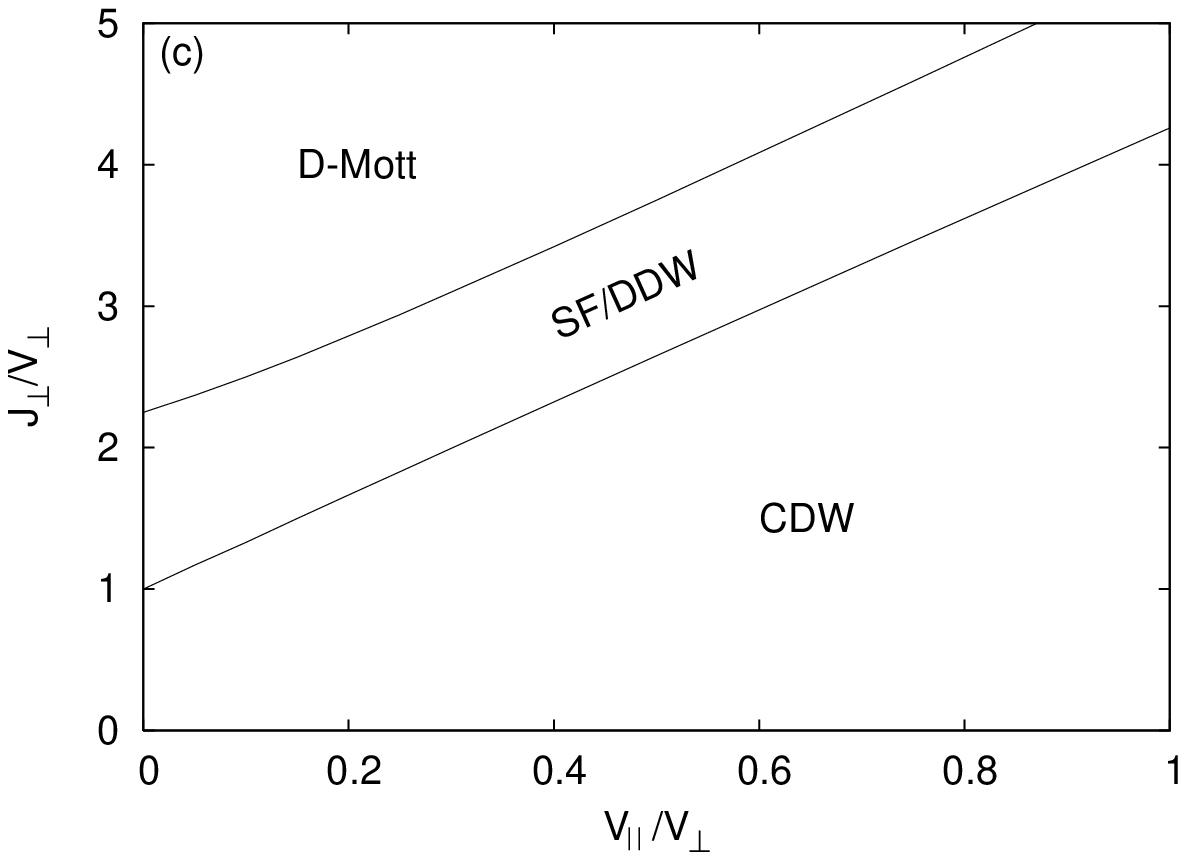}}
\caption{Weak-coupling phase diagrams at half-filling, for
$V_{\perp}>0$. Of the four phases (illustrated in Fig.
\ref{picphases}) that appear in these phase diagrams, the SF/DDW and
CDW phase have true long-range order. The SF/DDW phase appears
between the CDW and D-Mott phase. (a) $J_{\parallel}=V_{\parallel}=0$.
The half-filled model has SO(5)
symmetry along the dashed line. The phase boundary line
$U+(3/4)J_{\perp}=V_{\perp}$ is independent of the filling and valid
for arbitrary interaction strengths, as shown in Refs.
\onlinecite{momhik03,momhik04}. (b) $V_{\parallel}=0$,
$U/V_{\perp}=0.25$. (c) $J_{\parallel}=0$,
$U/V_{\perp}=0.25$.} \label{phasediag}
\end{figure}

Fig. \ref{phasediag}(a) shows the phase diagram for
$J_{\parallel}=V_{\parallel}=0$ and $V_{\perp}>0$ in the
weak-coupling limit.\cite{schetal03} For the region $U>0$ we may
compare our results to those in Ref. \onlinecite{tsufur02} and find
excellent agreement. The SF/DDW phase appears between the D-Mott and
CDW phases and is seen to extend well beyond the region where it was
first realized to exist,\cite{fjamar02,marfjasud02} which was
characterized by exact or approximate SO(5)
symmetry\cite{scazhahan98,linbalfis98} (the SO(5) symmetry occurs
along the dotted line in the figure). The nature of the quantum
phase transitions between the various phases has been discussed in
detail in Refs. \onlinecite{tsufur02,wuliufra03} (see also Ref.
\onlinecite{linbalfis98}).

Next, we consider the effects of including interactions along the
legs. We fix $U/V_{\perp}=0.25$ and add a finite $J_{\parallel}/V_{\perp}$
(Fig. \ref{phasediag}(b)) or $V_{\parallel}/V_{\perp}$
(Fig. \ref{phasediag}(c)). $J_{\parallel}/V_{\perp}$ is seen to
favor the D-Mott phase. This squeezes the SF/DDW phase which
eventually disappears for large enough values of
$J_{\parallel}/V_{\perp}$. On the other hand,
$V_{\parallel}/V_{\perp}$ is seen to favor the CDW phase, which is
easy to understand intuitively. However, in this case the SF/DDW
order is more robust: the width of the SF/DDW region remains more or
less constant as $V_{\parallel}/V_{\perp}$ is increased, while its
location is pushed towards larger values of $J_{\perp}/V_{\perp}$.

From Fig. \ref{phasediag} the conditions that favor SF/DDW order for
repulsive interactions and positive exchange constants can be
roughly summarized as follows:\cite{tsufur02} the ratios
$J_{\perp}/U$ and $V_{\perp}/U$ must be sufficiently large, but not
too large (as that favors the D-Mott and CDW phase respectively),
with leg interactions $J_{\parallel}$ and $V_{\parallel}$ preferably
zero or at least small compared to the other interaction parameters.
One might speculate that the reason why anisotropic interactions are
required to stabilize SF/DDW order in the two-leg ladder could be
related to the fact that the currents in the SF/DDW phase must
themselves be anisotropic in this geometry in order to satisfy
current conservation (the rung currents are twice as large as the
leg currents). It is also interesting to note that in
mean-field\cite{nervac89,norole04} and
renormalization-group\cite{binbaedou02,kamkat03} studies of
generalized Hubbard models on the square lattice in two dimensions,
at or near half-filling, and with \textit{isotropic} interactions
(i.e., $J_{\perp}=J_{\parallel}=J$ and $V_{\perp}=V_{\parallel}=V$),
the conditions that are most favorable for SF/DDW ordering
tendencies also appear to be that the ratios $J/U$ and $4V/U$ be
sufficiently large.

\section{DMRG study of the doped ladder}
\label{dmrg}

In this section we discuss an extensive
DMRG\cite{whi9293,peswankauhal98,sch04} study of the ground state of
hole-doped two-leg ladders with up to 200 rungs for intermediate
interaction strengths. The doping away from half-filling is measured
by the parameter $\delta$, defined as
\begin{equation}
\delta \equiv 1-n = \frac{{\cal N}_h}{2L},
\end{equation}
where $n$ is the average number of electrons per site, and ${\cal
N}_h$ and $L$ are the number of (doped) holes and rungs,
respectively. Open boundary conditions were used, as these are
computationally preferable in DMRG. The weak-coupling phase diagrams
at half-filling discussed in Sec. \ref{phasediagrams} were used as guides to suggest
interesting parameter regimes.

\begin{figure}
\centerline{\includegraphics[scale=0.6]{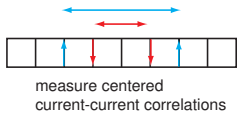}}
\caption{(Color online) In real-valued DMRG, possible long-ranged current
patterns are detected by calculating current-current correlations
centered around the ladder middle.} \label{fig:real_DMRG}
\end{figure}

To detect the existence of orbital currents, we have considered two approaches:

(i) Within a standard DMRG calculation in real number space, we have
studied the decay of rung-rung correlations where the rungs have
been centered about the middle of the ladder to minimize boundary
effects (Fig. \ref{fig:real_DMRG}):
\begin{equation}
C(r)=\langle j_{\perp}(L/2+r/2) j_{\perp}(L/2-r/2) \rangle ,
\end{equation}
where the rung current $j_{\perp}(x)={\cal J}_{(t,x),(b,x)}$ and the
current operator between two nearest-neighbor sites ${\bf i}$ and
${\bf j}$ is given by
\begin{equation}
{\cal J}_{{\bf i},{\bf j}} = i t \sum_{s} \left( c^\dag_{{\bf
i},s}c_{{\bf j},s} - c^\dag_{{\bf j},s}c_{{\bf i},s}\right).
\end{equation}

(ii) Within a DMRG calculation generalized to complex number space,
one may obtain non-vanishing current expectation values,
\begin{equation}
    \langle {\cal J}_{{\bf i},{\bf j}} \rangle \neq 0.
\end{equation}
\begin{figure}
\centerline{\includegraphics[scale=0.6]{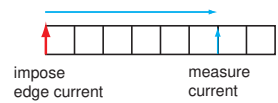}}
\caption{(Color online) In complex-valued DMRG, possible long-ranged
current patterns are detected by introducing a symmetry-breaking
infinitesimal surface current and observing whether it sustains a
long-ranged current pattern.} \label{fig:complex_DMRG}
\end{figure}
In the textbook case of a ferromagnet that develops a finite
magnetization while spontaneously breaking the continuous rotational
symmetry, this is achieved in theoretical approaches by forcing a
possible symmetry breaking by introducing an infinitesimally strong
external field.\cite{marston} In close analogy, we break the $Z_{2}$
symmetry of the direction of plaquette currents by introducing an
infinitesimal surface current $-h j_{\perp}(1)$ on the first rung of
the ladder, corresponding to a weak surface field (Fig.
\ref{fig:complex_DMRG}). To assure that this infinitesimal surface
(or edge) current does not change the physics of the ladder under
study, we have taken it down to values of $0.0001t$ (in some special
cases even lower values were checked). We find that results do not
change qualitatively while decreasing the surface current: in the
phase that does support SF/DDW order, the magnitude of the orbital
current measured numerically does not depend on its value in the
small current limit, while in the phases that do not support SF/DDW
order, the current decays exponentially fast into the bulk and its
magnitude is proportional to the applied surface current.

\begin{figure}
\centerline{\includegraphics[scale=0.5]{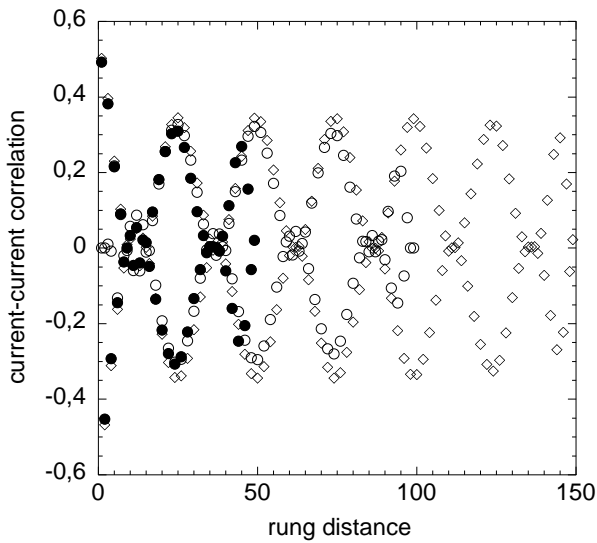}} \caption{Rung
current-current correlations $C(r)$ calculated for ladders of length
50 (solid circles), 100 (open circles), 150 (diamonds) in the SF/DDW
phase for $U=-0.4$, $t=1$, $V_{\perp}=0.9$, $J_{\perp}=2$, and
doping $\delta=0.08$. $J_{\parallel}=V_{\parallel}=0$ in this and
all other figures in this section.} \label{fig:correl8}
\end{figure}

\begin{figure}
\centerline{\includegraphics[scale=0.5]{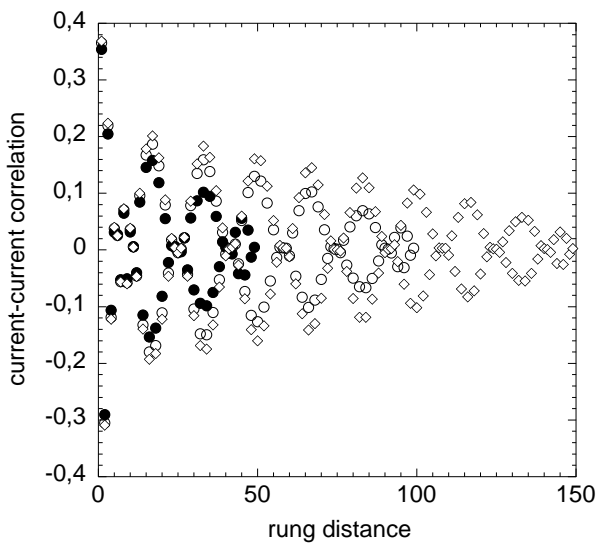}} \caption{Rung
current-current correlations $C(r)$ as in Figure \ref{fig:correl8},
but doping $\delta=0.12$.} \label{fig:correl12}
\end{figure}

For a large number of cases, we have performed both calculations and
found results in very good agreement. However, the first approach,
while computationally much less expensive, shows, due to the open
boundary conditions used in the DMRG runs, much more pronounced
finite-size effects: SF/DDW correlations are suppressed, not
enhanced by the presence of open boundary conditions. Long-ranged or
quasi-long-ranged order is thus often much harder to detect. Figs.
\ref{fig:correl8} and \ref{fig:correl12} show the current-current
correlations for a point in parameter space which exhibits SF/DDW
order both for 8 and 12 percent doping and various ladder lengths.
In the first case, correlations are weakly suppressed by the edges
and order is demonstrated in agreement with the second approach; in
the latter, correlations are strongly suppressed by the edges and
grow strongly in size with ladder length; even for 200 rungs, it
would be impossible to detect long-ranged order which is clearly
visible from 50 rungs upwards in the second approach. For parameter
sets where accessible system sizes allow to see long-ranged
correlations also in the rung-rung current correlations, the values
of the rung currents that can be extracted are in very good
agreement with the values found by imposing an infinitesimal edge
current.

\begin{figure}
\centerline{\includegraphics[scale=0.5]{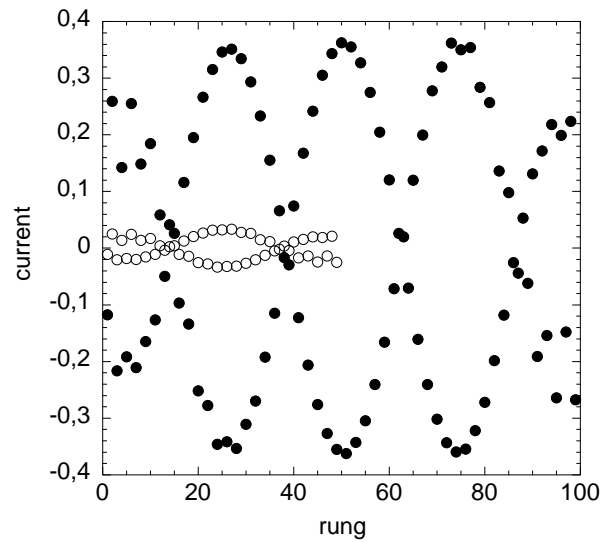}}
\caption{Rung current at the phase boundary between SF/DDW and CDW
order. The edge current is $0.0001t$ and $U=0.25$, $t=V_{\perp}=1$,
$J_{\perp}=1$, and doping $\delta=0.04$ for ladder lengths 50 (open
circles) and 100 (solid circles). } \label{fig:cdw_vs_ddw}
\end{figure}

In the second, computationally much harder, approach, we also
observe that on very short ladders right at the CDW-SF/DDW phase
boundary the open boundary conditions strongly favor a checkerboard
charge density wave over the SF/DDW currents (Fig.
\ref{fig:cdw_vs_ddw}), but the magnitude of the plaquette currents
stabilizes very quickly with increasing system length, such that we
have preferred this approach in our numerical calculations.  In Refs.
\onlinecite{momhik03,momhik04} exact duality relations were used to ascertain
that the phase boundary between the SF/DDW and CDW phases lies at $U
- V_{\perp} + (3/4) J_{\perp} = 0$.  It is remarkable that, as
evident in Fig. \ref{fig:cdw_vs_ddw}, strong currents are found to
persist right at the phase boundary separating the SF/DDW and CDW
phases. In this particular calculation, we have taken the edge
current as small as $10^{-8}t$ without observing any change in the
results.

\begin{figure}
\centerline{\includegraphics[scale=0.5]{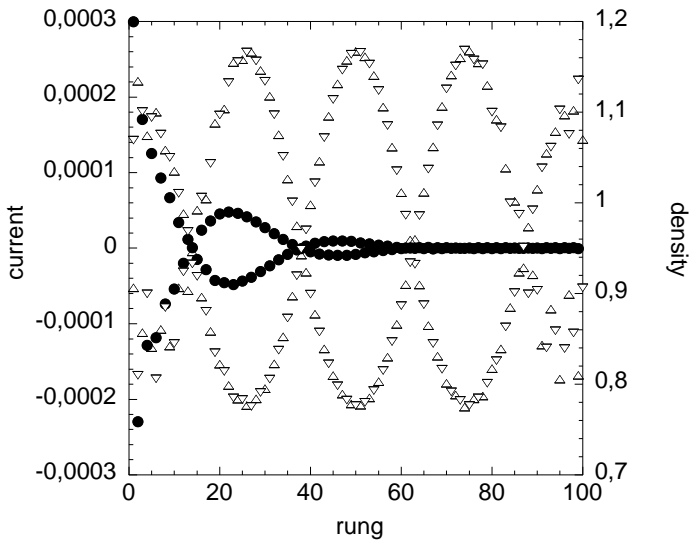}} \caption{Rung
current (solid circles, left $y$-axis, units of $t$) and electronic
densities on top and bottom ladder leg (up and down triangles, right
$y$-axis) for 100 rung-ladder in the CDW phase and edge current
$0.0001t$ ($U=0.25$, $t=V_{\perp}=1$, $J_{\perp}=0.8$, and doping
$\delta=0.04$).} \label{fig:cdw_type}
\end{figure}

\begin{figure}
\centerline{\includegraphics[scale=0.5]{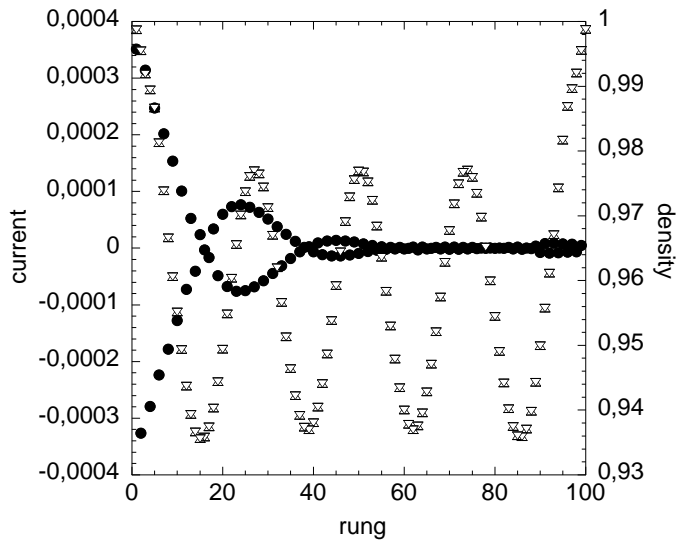}}
\caption{Rung current (solid circles, left $y$-axis, units of $t$)
and electronic densities on top and bottom ladder leg (up and down
triangles, right $y$-axis) for 100 rung-ladder in the doped D-Mott
phase and edge current $0.0001t$ ($U=0.25$, $t=V_{\perp}=1$,
$J_{\perp}=1.7$, and doping $\delta=0.04$). Top and bottom densities
coincide. Note that for these model parameters there is SF/DDW order
at half-filling (see Fig. \ref{phasediag}(a)); the phase boundary
between the D-Mott and SF/DDW phase moves downwards when the system
is doped.} \label{fig:dmott_type}
\end{figure}

\begin{figure}
\centerline{\includegraphics[scale=0.5]{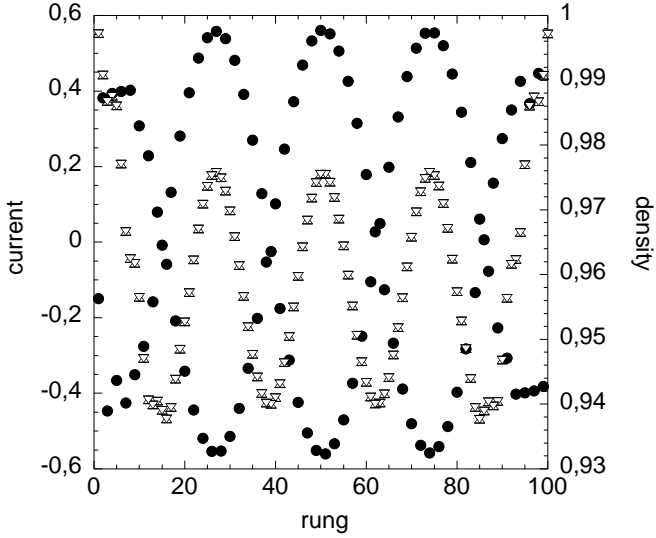}} \caption{Rung
current (solid circles, left $y$-axis, units of $t$) and electronic
densities on top and bottom ladder leg (up and down triangles, right
$y$-axis) for 100 rung-ladder in the SF/DDW phase and edge current
$0.0001t$ ($U=0.25$, $t=V_{\perp}=1$, $J_{\perp}=1.5$, and doping
$\delta=0.04$). Top and bottom densities coincide.}
\label{fig:ddw_type}
\end{figure}

The SF/DDW phase can be clearly distinguished from the neighboring
CDW and doped D-Mott phases.\cite{DMottdoped} In both the CDW (Fig.
\ref{fig:cdw_type}) and doped D-Mott (Fig. \ref{fig:dmott_type})
phase, the edge current induces rung currents $j_{\perp}(x)$ that
die out exponentially fast, while in the SF/DDW phase it stabilizes
a rung current orders of magnitude larger (Fig. \ref{fig:ddw_type}).
In the SF/DDW phase the {\em staggered} rung current $j_s(x)$
(defined in Eq. (\ref{js})) oscillates around zero with a
characteristic wavelength $2/\delta$ (this wavelength tends to
decrease slightly with decreasing distance to the boundary). Thus
the rung currents have anti-phase domain walls separated by a
characteristic distance $1/\delta$. At these domain walls (or
``discommensurations") the rung current has the same sign on
adjacent rungs ($++$ or $--$) instead of alternating between rungs
as in the (``commensurate") regions between domain walls.
Furthermore, using that
\begin{equation}
2k_F\equiv k_{F1}+k_{F2}=\pi n=\pi(1-\delta),
\label{2kF}
\end{equation}
which is the generalization of Eq. (\ref{kFsum}) away from
half-filling, one observes that the rung current oscillations may be
associated with the fundamental wavevector $2k_F$ as these
oscillations vary approximately as $\cos 2k_F x=(-1)^x \cos
\pi\delta x$ (this equality holds since the rung index $x$ is an
integer). Moreover, the currents on the top and bottom leg are
always opposite in sign. Consequently, the ordering wavevector of
the currents is $(2k_F,\pi)$, which reduces to $(\pi,\pi)$ at
half-filling, as expected.

A modulation of the rung charge density coexists with the orbital
currents in the SF/DDW phase (Fig. \ref{fig:ddw_type}). The
characteristic wavelength of this modulation is $1/\delta$ (again,
the wavelength decreases slightly as the distance to the boundary
decreases), which is just the average distance $L/({\cal N}_h/2)$
between \textit{pairs} of doped holes. Thus there are two doped
holes per wavelength, one per leg; in a cartoon picture the doped
hole pairs simply distribute themselves equidistantly along the
ladder, each hole pair occupying one rung. These hole
pairs tend to stay in the vicinity of the rung current minima
(the domain walls of the current pattern). The ordering wavevector
for this charge density modulation is $(4k_F,0)$ because its
characteristic periodicity in the direction parallel to the legs is
the same as that of $\cos 4k_F x = \cos 2\pi\delta x$, and the
density is the same on the top and bottom site of a rung. Clearly,
this charge density modulation disappears at half-filling.

As seen in Fig. \ref{fig:dmott_type}, the $(4k_F,0)$ modulation of
the charge density exists also in the doped D-Mott phase. The nature
of the charge density pattern in the CDW phase (Fig.
\ref{fig:cdw_type}) is less obvious at first sight; however, as will
become apparent in Sec. \ref{bosdoped}, it is the sum of a large
component with wavevector $(2k_F,\pi)$ and a smaller component with
wavevector $(4k_F,0)$. Thus the $(4k_F,0)$ charge density modulation
is present in all three phases appearing in the parameter region
considered here. Such $(4k_F,0)$ charge density modulations in
two-leg ladders have recently been discussed by White {\em et
al.};\cite{whiaffsca02} we will analyze these modulations in more
detail in Sec. \ref{bosdoped}.

\begin{figure}
\centerline{\includegraphics[scale=0.75]{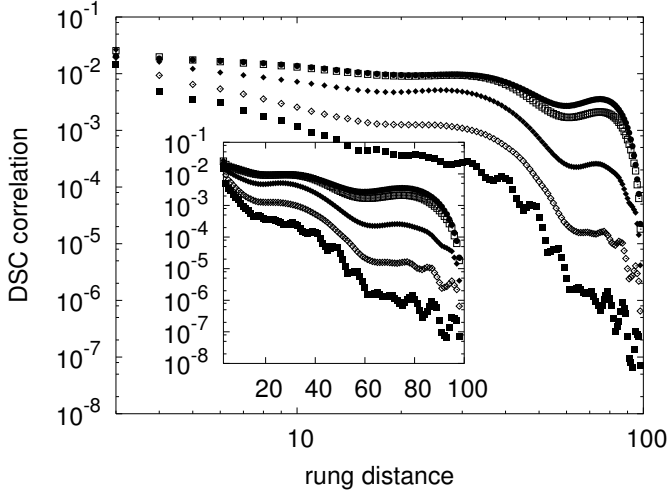}} \caption{Log-log
plot of DSC correlations for a 100-rung ladder with $U=0.25$,
$t=V_{\perp}=1$ and doping $\delta=0.04$, for different values of
$J_{\perp}$ (from bottom to top): 1.3 (solid squares), 1.5 (open
diamonds), 1.7 (solid diamonds), 2.0 (open squares), and 2.6 (solid
circles). The first two values are in the SF/DDW phase, the last
three in the doped D-Mott phase. The inset shows a lin-log plot of the same
data.} \label{fig:dsc}
\end{figure}

We have also calculated the DSC correlations in the SF/DDW and doped
D-Mott phases (Fig. \ref{fig:dsc}). In the SF/DDW phase the DSC
correlations decay exponentially. As expected, the DSC correlations
fall off more slowly in the doped D-Mott phase. Close to the phase
boundary to the SF/DDW phase the correlations appear still to be
exponential, however, while further into the doped D-Mott phase the
edge effects make it difficult to say whether for longer system
sizes the correlations might turn out to be algebraic.

Most of our DMRG calculations are for particle densities close to
half-filling. As DMRG is a canonical ensemble method, particle
numbers are fixed at each growth step and there is some
arbitrariness in inserting the particles and holes to maintain
essentially constant particle density during system growth. Various
insertion schemes have yielded essentially identical results. To
obtain such well-converged results, it is crucial to apply several
runs (sweeps) of the finite-system DMRG algorithm until convergence
is observed numerically. We have typically performed of the order of
10 to 20 sweeps and observed convergence after typically 5 to 7
sweeps.

\begin{figure}
\centerline{\includegraphics[scale=0.5]{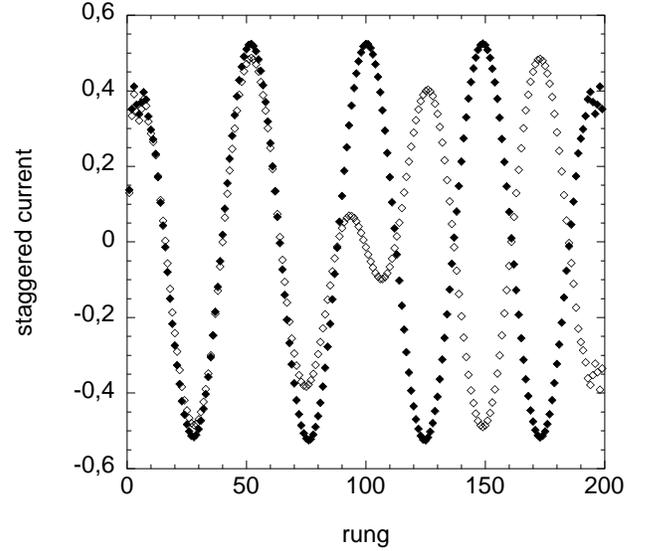}}
\caption{Staggered rung current for 200 rung-ladder in the SF/DDW
phase and edge current $0.0001t$ ($U=0.25$, $t=V_{\perp}=1$,
$J_{\perp}=1.5$, and doping $\delta=0.04$). Final number of states
$m=400$ was reached starting from $m=100$ (open diamonds) and
$m=400$ (solid diamonds).} \label{fig:statehistory}
\end{figure}

Up to 800 states were kept for DMRG runs; following standard
practice, the number of states kept was initially chosen to be much
smaller and increased during sweeps. For systems up to about 100
rungs, final results were independent of the way the number of
states was augmented. For systems well beyond 100 rungs we have
observed that in the SF/DDW phase, DMRG runs that started with a
very low number of states kept, converged to a result where the
current amplitudes were somewhat suppressed with respect to runs
that started with high precision and yielded results in perfect
agreement with those found for shorter systems. The low precision
calculation introduces a phase slip $\pi$ at the center (Fig.
\ref{fig:statehistory}).

\begin{figure}
\centerline{\includegraphics[scale=0.5]{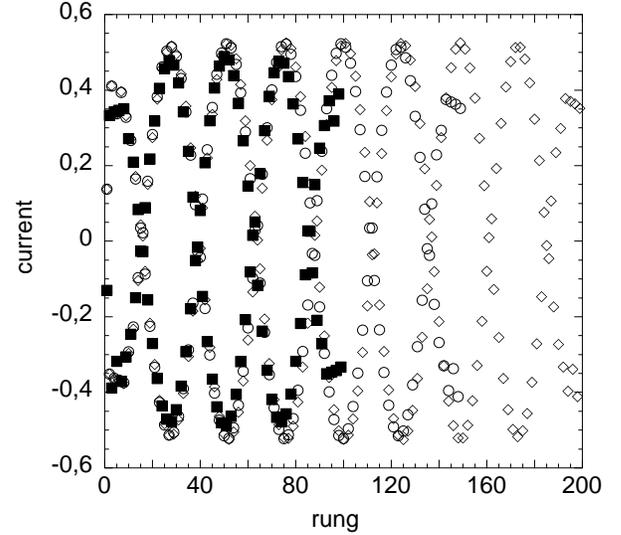}}
\caption{Rung current in the SF/DDW phase and edge current $0.0001t$
($U=0.25$, $t=V_{\perp}=1$, $J_{\perp}=1.5$, and doping
$\delta=0.04$) for ladder lengths 100 (solid squares), 150
(circles), 200 (diamonds).} \label{fig:correls100to200}
\end{figure}

\begin{figure}
\centerline{\includegraphics[scale=0.5]{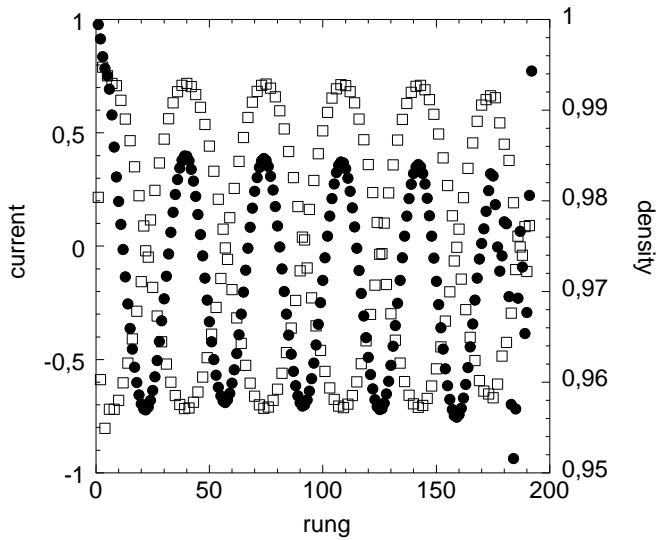}} \caption{Rung
current (open squares, left $y$-axis, units of $t$) and electronic
densities (solid circles, right $y$-axis) for 192-rung ladder in the
SF/DDW phase ($U=-0.4$, $t=1$, $V_{\perp}=0.9$, $J_{\perp}=2$), edge
current $0.0001t$ and 11 holes (i.e., $\delta\approx 0.0286$,
$1/\delta\approx 34.9$). The disturbed density pattern and slight
current reduction on the far end of the ladder is due to the
presence to one unpaired hole.} \label{fig:holes11}
\end{figure}

Do we really have true long-range order in the orbital currents? If
there is not true long-range order, one might speculate that the
introduction of a phase slip (of the type seen in Fig.
\ref{fig:statehistory}) at low energetic cost is the way how
long-ranged correlations are lost in this system. However, we did
not observe, for systems up to 200 rungs, any significant decrease
of the magnitude of the triggered current pattern. Considering Fig.
\ref{fig:correls100to200}, one actually sees that maximum current
amplitudes grow from system size 50 to 100, in line with the SF/DDW
suppression by edges reported above, to stay constant after that.
One might envisage that the open boundary conditions lock in the
currents, leaving us with seeming long-ranged order. The repeated
observation that open boundary conditions seem to disfavor SF/DDW
does not support this point of view. It might also be speculated
that our filling factors, 4, 8 and 12 percent away from
half-filling, allow the formation of commensurate current patterns
which may show long-ranged order as opposed to the generic case of
incommensurate filling. Introducing 11 holes on a 192 rung ladder to
model an approximation to a generic incommensurate filling on a
finite system, however, we do not observe current decay either (Fig.
\ref{fig:holes11}).

\begin{figure}
\centerline{\includegraphics[scale=0.5]{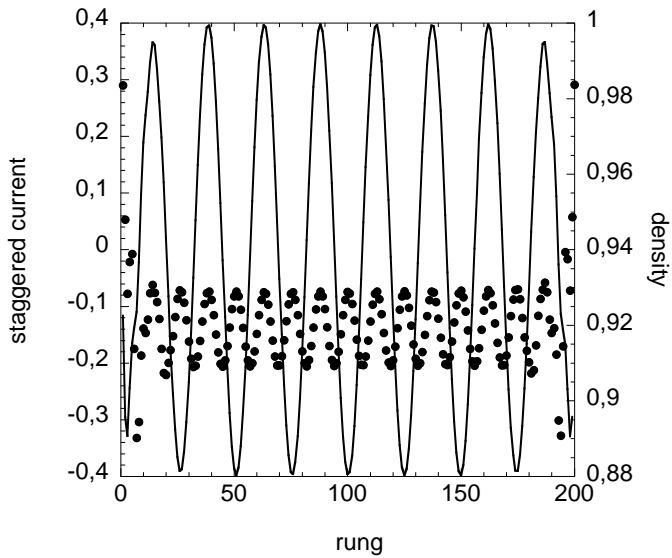}}
\caption{Rung current (lines, left $y$-axis, units of $t$) and
electronic densities (solid circles, right $y$-axis) for 200-rung
rung-ladder in the SF/DDW phase for $\delta=0.08$ and edge current
$0.0001t$ ($U=0.25$, $t=V_{\perp}=1$, $J_{\perp}=1.5$).}
\label{fig:new8percent}
\end{figure}

The SF/DDW phase survives up to quite substantial dopings away from
half-filling. Fig. \ref{fig:new8percent} shows current and density
oscillations in the SF/DDW phase of a 200-rung ladder for doping
$\delta=0.08$. For the parameter sets studied, the phase transition
where SF/DDW order is finally suppressed to zero seems to occur
roughly between 10 and 20 percent doping; a more precise estimate
would require a more systematic study of the doping dependence of the
SF/DDW phase as a function of the interaction parameters.

\section{Effective field theory description of the doped ladder}
\label{bosdoped}

In this section we try to understand the DMRG results in Sec.
\ref{dmrg} using weak-coupling RG/bosonization arguments.
We will be particularly interested in what these arguments
predict regarding the possibility of having true long-range
order for the considered doping levels.

\subsection{Physical observables}
\label{physobsdoped}

We first present the set of relevant physical observables and their
bosonized expressions for the case of an arbitrary filling for which
the Fermi energy crosses both bands. This includes the half-filled
case discussed in Sec. \ref{weakint}, as well as the doping levels
studied numerically in Sec. \ref{dmrg}.

The rung current operator is
\begin{eqnarray}
j_{\perp}(x) &=& j_{2k_F} \big[\cos(2k_F x+\phi_{+\rho})\cos \phi_{+\sigma}
\cos \theta_{-\rho}\cos \theta_{-\sigma} \nonumber \\ & & \hspace{-1.0cm}
+\;\sin(2k_F x+\phi_{+\rho})\sin \phi_{+\sigma}\sin\theta_{-\rho}\sin\theta_{-\sigma}\big].
\label{rungcurarb}
\end{eqnarray}
Furthermore, the density operator $n_{\ell}(x)$ on leg $\ell$ is
\begin{eqnarray}
\lefteqn{
n_{t,b}(x) = (1-\delta) + \frac{1}{\pi}\frac{\partial\phi_{+\rho}}{\partial x}} \nonumber \\
 &\pm & n_{2k_F}\big[\cos(2k_F x+\phi_{+\rho})\cos\phi_{+\sigma}\sin\theta_{-\rho}\cos\theta_{-\sigma} \nonumber \\
  &-& \sin(2k_F x+\phi_{+\rho})\sin\phi_{+\sigma}\cos\theta_{-\rho}\sin\theta_{-\sigma}\big] \nonumber \\
&+& n_{4k_F}\cos(4k_F x+2\phi_{+\rho})\cos 2\phi_{+\sigma}.
\label{densityarb}
\end{eqnarray}
The quantity $2k_F$ was defined in Eq. (\ref{2kF}). The nonuniversal
coefficients $j_{2k_F}$, $n_{2k_F}$ and $n_{4k_F}$ depend on the
short-distance cutoff of the theory.

The $4k_F$ term in $n_{\ell}(x)$ should be particularly noted. It
does not come out of a naive calculation of the density operator
using the bosonization formula (\ref{bosfieldop}); however, a more
general, phenomenological reasoning shows that such higher harmonics
terms are generally expected.\cite{halprl81,gia03} The form of the
$4k_F$ term in $n_{\ell}(x)$ was deduced by White {\em et
al.}\cite{whiaffsca02} They considered the $4k_F$ term in the
correlation function for the {\em square} of the density operator,
as resulting from the product of the $2k_F$ terms in the density
operator calculated from the bosonization formula, and then
(implicitly) argued that a similar $4k_F$ contribution to that
correlation function would be expected to come from the product of
the constant term and a $4k_F$ term in the density operator. Higher
harmonics than $2k_F$ in $j_{\perp}(x)$ and $4k_F$ in $n_{\ell}(x)$
have been neglected in the expressions above as they are expected to
give at most only minor quantitative corrections to the terms
already included.

The density operator above is more complicated than the one
discussed for the half-filled CDW phase in Sec. \ref{OPs}. It is
therefore instructive to see how the expectation value of
$n_{\ell}(x)$ reduces to the correct form at half-filling. In that
case, both $\phi_{+\rho}$ and $\phi_{+\sigma}$ are locked in all
four phases (see Table \ref{phasetable}). This gives $\langle
\partial_x \phi_{+\rho}\rangle = 0$ and $\langle \cos
2\phi_{+\sigma}\rangle \neq 0$; furthermore, from $4k_F=2\pi$ it
follows that $\langle\cos(4k_F x+2\phi_{+\rho})\rangle = \langle
\cos 2\phi_{+\rho}\rangle \neq 0$. Thus the expectation value of the
$4k_F$ term is independent of $x$ at half-filling. Consequently, the
only way to obtain the correct average density of one electron per
site is to have $n_{4k_F}\equiv 0$ in this case. Hence the
expectation value of the deviation of the density from its average
value, $\delta n_{\ell}(x)\equiv n_{\ell}(x)-1$, reduces to the
expectation value of the $2k_F$ term, which is only nonzero in the
CDW phase, in agreement with the simplified discussion of the
density operator in Sec. \ref{OPs}.

Finally, we note that since the DSC and SSC operators have zero
momentum, their bosonized expressions are the same as at half-filling;
see Eqs. (\ref{DSC-SSC-PO})-(\ref{SSC-OP}). (We will not consider SSC
order in the following, since the doped S-Mott phase was not studied
in Sec. \ref{dmrg}.)

\subsection{True long-range order scenario}
\label{tlroscen}

It was argued in Sec. \ref{dmrg} that the DMRG results are
consistent with true long-range orbital current and charge density
wave order. From the bosonization point of view, to have true
long-range order it is necessary that certain bosonic fields become
locked to appropriate values. In this subsection we show that if
this locking occurs, the resulting expectation values and
correlation functions of the physical observables of interest, as
calculated from the bosonized expressions in Sec.
\ref{physobsdoped}, are in good qualitative agreement with the DMRG
results. The assumption that the locking occurs is easily justified
for all bosonic fields in question except the symmetric charge field
$\phi_{+\rho}$. The conditions for locking this field will be
investigated in detail in Sec. \ref{condtrue}.

\begin{figure}
\centerline{\includegraphics[scale=0.75]{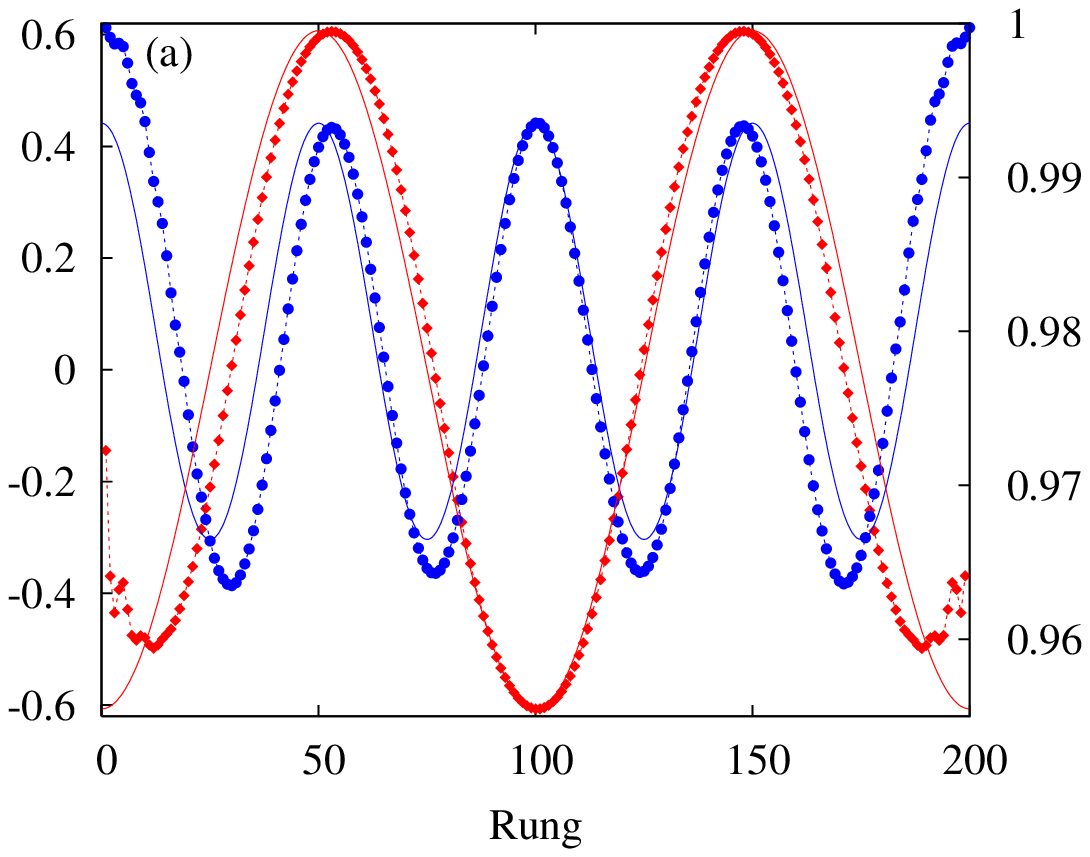}}
\centerline{\includegraphics[scale=0.75]{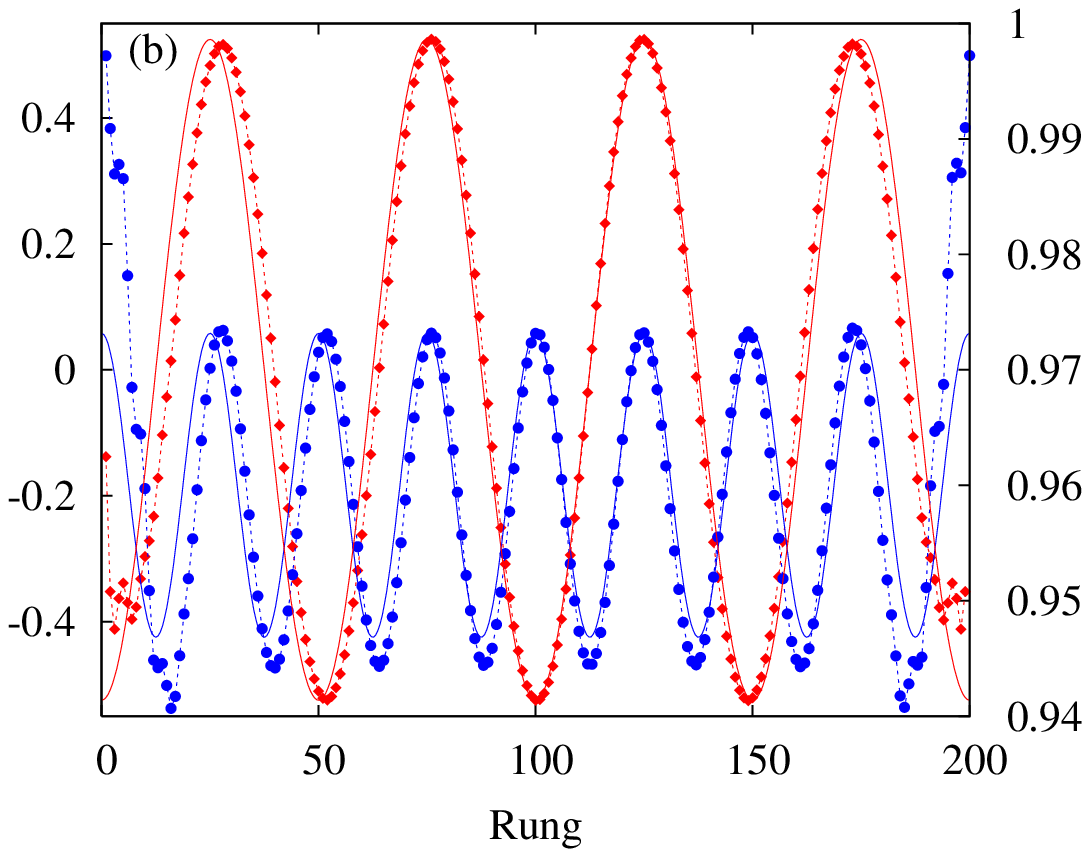}}
\caption{(Color online) (a) Staggered rung current $j_s(x)$ (red
curves, left y-axis) and electron density per site $n(x)$ (blue
curves, right y-axis) as a function of rung index $x$ in the SF/DDW
phase of a 200-rung ladder with $\delta=0.02$. Shown are DMRG
results (diamonds and circles for currents and densities,
respectively, connected by dashed lines to guide the eye) and
analytical fits from the true long-range order scenario (full
lines). The model parameters for the DMRG calculation are
$t=V_{\perp}=1$, $U=0.25$, $J_{\perp}=1.5$ and
$J_{\parallel}=V_{\parallel}=0$. The values of the fitting
parameters are $|j_0|\approx 0.61$ and $n_0\approx 0.0135$. (b)
Plots for the same model parameters as above, except that the doping
$\delta=0.04$ and the values of the fitting parameters are
$|j_0|\approx 0.52$ and $n_0\approx 0.0132$. }
\label{fig:trueLROfit}
\end{figure}

Let us first consider the SF/DDW phase. True long-range order in the
currents, $\langle j_{\perp}(x)\rangle \neq 0$, requires that the
four fields $\phi_{+\rho}$, $\phi_{+\sigma}$, $\theta_{-\rho}$, and
$\theta_{-\sigma}$ become locked to appropriate values as determined
by Eq. (\ref{rungcurarb}). The momentum-conserving interaction
(\ref{HI1b}), which contains cosines of the three latter fields, is
present also away from half-filling, and weak-coupling one-loop RG
calculations\cite{balfis96,sch96,origia97,wuliufra03} show that it
can cause $\phi_{+\sigma}$, $\theta_{-\rho}$, and $\theta_{-\sigma}$
(or $\phi_{-\sigma}$) to lock also in the doped case, at least if
the doping is not too large. This picture is expected to hold also
for stronger interactions. Let us now assume that also the symmetric
charge field $\phi_{+\rho}$ can become locked for the dopings
considered in Sec. \ref{dmrg}. There is then a finite energy gap to
excitations in $\phi_{+\rho}$. Using the field-locking pattern for
$\phi_{+\sigma}$, $\theta_{-\rho}$, and $\theta_{-\sigma}$ in Table
\ref{phasetable}, and evaluating the expectation value of the
current operator semiclassically,\cite{semiclapp} we find
\begin{equation}
\langle j_{\perp}(x)\rangle \approx j_0\cos(2k_F x + \langle
\phi_{+\rho}\rangle), \label{j0}
\end{equation}
with $j_0=j_{2k_F}\langle \cos \phi_{+\sigma}\cos \theta_{-\rho}\cos
\theta_{-\sigma}\rangle\neq 0$. Since $\phi_{+\rho}$ and
$\phi_{+\sigma}$ are locked, we also have (with $n_t(x)=n_b(x)\equiv
n(x)$ in the SF/DDW phase)
\begin{equation}
\langle n(x)\rangle = (1-\delta)+n_0\cos(4k_F x + 2\langle
\phi_{+\rho}\rangle), \label{n0}
\end{equation}
with $n_0=n_{4k_F}\langle \cos 2\phi_{+\sigma}\rangle\neq 0.$ This
shows that $(2k_F,\pi)$-SF/DDW order and $(4k_F,0)$-CDW order
coexist in the SF/DDW phase. Furthermore, if the non-universal
coefficient $n_{4k_F}$ is positive, the minima of the electron
density always occur at the zeros (anti-phase domain walls) of the
current. These results are in agreement with the findings in Sec.
\ref{dmrg}. The different solutions for $\langle
\phi_{+\rho}\rangle$ give rise to degenerate ground states which are
related to each other by translations by a lattice vector (in the
DMRG calculations, this degeneracy is of course lifted by the open
boundary conditions, leaving only a two-fold degeneracy due to the
breaking of time reversal symmetry). Fig. \ref{fig:trueLROfit} shows
a fit of these analytical results for the SF/DDW phase to the DMRG
results for two different dopings. The agreement is quite good. If
the true long-range order scenario is in fact realized, we expect
that the minor differences between the numerical and analytical
results could be removed, at least in principle, by improving the
latter by taking into account boundary and finite-size effects,
corrections due to the presence of higher harmonics, and by going
beyond the semiclassical approximations used in evaluating the
expectation values.

By using similar arguments as for the SF/DDW phase, one also
predicts, again in agreement with DMRG results, that there is
coexistence of $(2k_F,\pi)$-CDW and $(4k_F,0)$-CDW order in the CDW
phase, while in the doped D-Mott phase there is ``only" $(4k_F,0)$-CDW
order. The reason for the ubiquity of $(4k_F,0)$-CDW order is that
this order parameter only contains the fields $\phi_{+\rho}$ and
$\phi_{+\sigma}$, both of which are locked in all phases in this
true long-range order scenario.

This scenario furthermore predicts that the asymptotic decay of DSC
correlations is exponential in all phases. The DSC correlation
length is expected to be smallest in the CDW phase because then the
field-locking differs from what the DSC operator wants in both the
$-\rho$, $-\sigma$, and $+\rho$ sectors. In the SF/DDW phase the
correlation length should be larger because there is no longer
disagreement in the $-\rho$ sector. In the doped D-Mott phase the
correlation length is larger still because the $-\sigma$ sector also
matches up, the only exponential decay left coming from the $+\rho$
sector. These predictions for the qualitative behavior of the DSC
correlations are in good agreement with the DMRG results (with the
qualifier that in the case of the doped D-Mott phase, the nature
of the asymptotic decay of DSC correlations is difficult to deduce
from the DMRG data; see the discussion in Sec. \ref{dmrg}).

\subsection{True long-range order and Umklapp interactions}
\label{condtrue}

The crucial assumption which enabled the true long-range order
scenario just discussed to emerge from the bosonization formulas was
that the symmetric charge field $\phi_{+\rho}$ can become locked for
the dopings and interaction parameters considered in Sec.
\ref{dmrg}. In this subsection we use symmetry arguments to derive
the form of the terms that, under the right circumstances, are able
to lock $\phi_{+\rho}$. It will be shown that these terms are
Umklapp interactions, and that they are allowed by symmetry in the
low-energy effective Hamiltonian only when the average number of
electrons per site, $n$, is a rational number $p/q$. The further
analysis suggests that the Umklapp interactions are only able to
lock $\phi_{+\rho}$ for relatively small values of $q$. Therefore
this weak-coupling RG/bosonization analysis appears not to support a
true long-range order scenario for the dopings considered in Sec.
\ref{dmrg}, which correspond to quite large ($\geq 25$) values of
$q$.

If the low-energy effective Hamiltonian, denoted by
$H_{\text{eff}}$, is invariant under global translations in
$\phi_{+\rho}$,
\begin{equation}
\phi_{+\rho}(x)\to \phi_{+\rho}(x)+\alpha\quad (\alpha \mbox{
arbitrary real)}, \label{phitransl}
\end{equation}
locking $\phi_{+\rho}$ implies that this continuous symmetry must be
spontaneously broken, which is impossible in one dimension.
Therefore, to lock $\phi_{+\rho}$ it is necessary that there be
terms in $H_{\text{eff}}$ that break this continuous symmetry.
Furthermore, the Hamiltonian can only contain gauge-invariant terms.
The bosonic fields themselves are not gauge-invariant, only
derivatives and exponentials of them are. Of these, only
exponentials of $\phi_{+\rho}$ are not invariant under
(\ref{phitransl}), as they break the continuous symmetry down to a
discrete one. Thus one is led to consider terms of the form
\begin{equation}
H_Q=F_Q+F_Q^{\dagger},\quad F_Q \equiv \int
dx\;A_Q(x)\,e^{iQ\phi_{+\rho}(x)},
\end{equation}
where $Q\neq 0$ and $A_Q(x)$ is an operator that we define to be
invariant under (\ref{phitransl}). We may take $Q>0$ without loss of
generality.

We will use the fact that $H_Q$ must be invariant under certain
symmetry operations (reflections and translations in real space) to
deduce under what conditions $H_Q$ can appear in $H_{\text{eff}}$,
and what the allowed values of $Q$ are in that case. However, to do
this it turns out that we will also need to know how $H_Q$
transforms under translations in $\phi_{-\rho}$. Since the conjugate
field $\theta_{-\rho}$ is locked in all phases, $\phi_{-\rho}$ is
disordered, so that an operator containing terms with exponentials
of $\phi_{-\rho}$ will be irrelevant in the RG sense. Thus, if $H_Q$
is to have a chance of causing true long-range order, we must assume
that $A_Q(x)$ does not contain such terms, which implies that $H_Q$
is invariant under global continuous translations in $\phi_{-\rho}$,
\begin{equation}
\phi_{-\rho}\to \phi_{-\rho}+\beta\quad (\beta \mbox{ arbitrary
real)}. \label{-rhotransl}
\end{equation}

Let us first consider reflection symmetry. $H_Q$ must be invariant
under the transformation
\begin{equation}
c_{t,x,s}\to c_{b,x,s},\quad c_{b,x,s}\to c_{t,x,s} \label{legint}
\end{equation}
which interchanges the ``top" and ``bottom" leg. By regarding the
two sites on a rung as forming a one-dimensional lattice in the
transverse ($y$) direction with periodic boundary conditions, we see
that (\ref{legint}) is a translation by one lattice spacing in this
two-site lattice, and is therefore effected by the unitary operator
$e^{-i{\cal P}_y}$ where ${\cal P}_y$ is the momentum operator in
the $y$ direction. The allowed transverse momenta are $k_y=0$ and
$\pi$, associated with the bonding and antibonding bands 2 and 1,
respectively. Thus
\begin{equation}
{\cal P}_y = \pi\sum_{x,s}c^{\dagger}_{1xs}c_{1xs} = \pi \sum_{P,s}
{\cal N}_{P1s}. \label{Pycont}
\end{equation}
One finds that $e^{-i{\cal P}_y}$ transforms the continuum field
operators as
\begin{equation}
\psi_{P1s}(x)\to -\psi_{P1s}(x),\quad \psi_{P2s}(x)\to
\psi_{P2s}(x). \label{12transf}
\end{equation}
Thus the symmetry under leg interchange implies that any term in the
Hamiltonian must contain an even number of both 1- and 2-operators
of this type (the latter due to conservation of the total number of
particles). Bosonization gives ${\cal P}_y = \int dx\; (\partial_x
\phi_{+\rho}-\partial_x \phi_{-\rho})$ which shows that $e^{-i{\cal
P}_y}$ generates translations in $\theta_{\pm \rho}$. We would
however like to find an operator that transforms the field operators
like (\ref{12transf}) while at the same time generating translations
in $\phi_{\pm \rho}$, as that would give information about $Q$. Such
an operator is $e^{-i{\cal R}}$, where $R$ is obtained by modifying
${\cal P}_y$ in (\ref{Pycont}) by a factor of $P$, which changes
$\phi\to\theta$ in the bosonic representation:
\begin{equation}
{\cal R} = \pi \sum_{P,s} P {\cal N}_{P1s} = \int dx\;(\partial_x
\theta_{+\rho}-\partial_x \theta_{-\rho}).
\end{equation}
As desired, $e^{-i{\cal R}}$ leaves all bosonic fields invariant
except $\phi_{\pm \rho}$ which transform as
\begin{equation}
\phi_{-\rho}\to \phi_{-\rho}-\pi,\quad \phi_{+\rho}\to
\phi_{+\rho}+\pi.
\end{equation}
Thus invariance of $H_Q$ under this transformation implies, when we
use that $H_Q$ must be invariant under (\ref{-rhotransl})
separately, that $Q$ must be an even integer.

Next, consider translation symmetry.\cite{aff90} $H_Q$ must be
invariant under the transformation
\begin{equation}
c_{\ell,x,s}\to c_{\ell,x+1,s}
\end{equation}
which translates the system by one lattice spacing in the $x$
direction. This transformation is generated by the operator
$e^{-i{\cal P}_x}$ where ${\cal P}_x$ is the momentum operator in
the $x$ direction, given by
\begin{eqnarray}
\lefteqn{{\cal P}_x = \sum_{k\lambda s}k\,n_{k\lambda s}\approx
\sum_{P\lambda s}Pk_{F\lambda}{\cal N}_{P\lambda s}} \nonumber \\
& & \hspace{-1.2cm}=\frac{1}{\pi}\!\int dx\Big[(k_{F1}+k_{F2})
\partial_x \theta_{+\rho}+(k_{F2}-k_{F1})\partial_x
\theta_{-\rho}\Big].
\end{eqnarray}
Clearly $e^{-i{\cal P}_x}$ leaves all bosonic fields unchanged
except $\phi_{\pm \rho}$ which transform as
\begin{eqnarray}
\phi_{-\rho} &\to & \phi_{-\rho}-k_{F1}+k_{F2}, \label{-rhotransf}\\
\phi_{+\rho} &\to & \phi_{+\rho}+k_{F1}+k_{F2} = \phi_{+\rho}+\pi n,
\label{+rhotransf}
\end{eqnarray}
where we used (\ref{2kF}) to express (\ref{+rhotransf}) in terms of
the average number of electrons per site, $n$. Again invoking the
invariance of $H_Q$ under (\ref{-rhotransl}), it follows that
(\ref{+rhotransf}) requires $Qn$ to be an even integer. This is
impossible if $n$ is an irrational number, implying that $H_Q$ is
not allowed in $H_{\text{eff}}$ in this case. Therefore true
long-range order is excluded for incommensurate fillings. If instead
$n$ is rational, i.e.,
\begin{equation}
n=p/q
\end{equation}
where $p$ and $q$ are coprimes (i.e., integers with no common
divisors), the values of $Q$ that make both $Q$ and $Qn$ even
integers are
\begin{equation}
Q=2rq,\quad r=1,2,\ldots.
\end{equation}
Therefore the terms $H_{2rq}$ are allowed in $H_{\text{eff}}$ in
this case.

We now consider the physical interpretation of these terms. As they
are not invariant under (\ref{phitransl}), they do not commute with
the generator of these translations, given by $G_{\phi_{+\rho}}=\int
dx\;\Pi_{\phi_{+\rho}} = ({\cal N}_L-{\cal N}_R)/2$, where
$\Pi_{\phi_{+\rho}}=-\partial_x \theta_{+\rho}/\pi$ is the conjugate
momentum of $\phi_{+\rho}$, and
\begin{equation}
{\cal N}_P\equiv \sum_{\lambda s}{\cal N}_{P\lambda s}=
\frac{1}{\pi}\int dx\;\partial_x(\phi_{+\rho}+P\theta_{+\rho})
\label{NP}
\end{equation}
is the number operator for right-/left-moving electrons. Because
$A_{2rq}(x)$ is invariant under independent global continuous
translations of $\phi_{+\rho}$ and $\theta_{+\rho}$ (the latter due
to conservation of total particle number), we have $[{\cal
N}_P,A_{2rq}(x)]=0$. This gives
\begin{equation}
[{\cal N}_P,F_{2rq}]=-2rq P \;F_{2rq}, \label{raise-lower}
\end{equation}
which shows that $F_{2rq}$ and $F_{2rq}^{\dagger}$ are
raising/lowering operators (by $2rq$ quanta) for ${\cal N}_P$:
$F_{2rq}$ scatters $2rq$ electrons from right to left while
$F^{\dagger}_{2rq}$ scatters them the other way. Furthermore, the
commutator
\begin{equation}
[{\cal P}_x,F_{2rq}]= - 4k_F rq\; F_{2rq}
\label{momtransf}
\end{equation}
shows that the associated momentum transfer is $\mp 4k_F rq$ for
$F_{2rq}$ and $F_{2rq}^{\dagger}$, respectively, which reduces to
$\mp 2\pi rp$ for $n=p/q$. Thus $H_{2rq}$ is an Umklapp interaction.
Note that the invariance of $H_Q$ under (\ref{-rhotransl}) was used
in the derivation of (\ref{momtransf}).

Of all the Umklapp interactions $H_{2rq}$, $r=1,2,\ldots$, that are
allowed in $H_{\text{eff}}$ for commensurate density $n=p/q$, the
operator $H_{2q}$ corresponding to $r=1$ is most relevant in the RG
sense, and is therefore the one we will focus on in the rest of the
discussion. $H_{2q}$ scatters $2q$ electrons from the left to the
right side of the Fermi surface (and vice versa), with associated
momentum transfer $\pm 2\pi p$.\cite{checkumklapp} The interaction
$H_{2q}$ with $q=1$ (which is allowed in $H_{\text{eff}}$ at
half-filling) will be referred to as the basic-Umklapp interaction.
The interactions $H_{2q}$ with $q>1$ will be referred to as
multiple-Umklapp interactions.

The derivation presented here shows that when the density $n$ is
moved away from the commensurate value $p/q$, the term $H_{2q}$ is
no longer allowed in $H_{\text{eff}}$. Another way of understanding
this is from the fact that when $n$ moves away from $p/q$,
$A_{2q}(x)$ acquires factors that oscillate in space (this is
intimately related to the fact that the momentum $4k_F q$
transferred by $H_{2q}$ now deviates from the reciprocal lattice
vector $2\pi p$), and these make $H_{2q}$ average to zero on
sufficiently long length scales, thereby rendering it an irrelevant
operator. For example, for doping $\delta$ away from half-filling
the basic-Umklapp interaction is proportional to
$\cos(2\phi_{+\rho}+4k_F x)=\cos(2\phi_{+\rho}+2\pi\delta x)$. Thus
the cosine oscillates with $x$ and averages to zero over length
scales much bigger than $1/\delta$.

Since the $-\rho$ and $\pm\sigma$ sectors are gapped out, the parts
of the Hamiltonian containing fields belonging to these sectors can
be replaced by their expectation values. The low-energy effective
Hamiltonian is then described in terms of the fields in the
remaining $+\rho$ sector. $H_{\rm{eff}}$ has a quadratic part given
by the Luttinger liquid Hamiltonian\cite{ph-breaking}
\begin{equation}
\hspace{-0.15cm}H_{\text{LL}}=\int
dx\,\frac{v_{+\rho}}{2\pi}\bigg[K_{+\rho}(\partial_x
\theta_{+\rho})^2 +\frac{1}{K_{+\rho}}(\partial_x
\phi_{+\rho})^2\bigg], \label{HLutt}
\end{equation}
where $v_{+\rho}$ and $K_{+\rho}$ are the effective velocity and
Luttinger parameters, respectively. $H_{\text{LL}}$ describes a
critical (Luttinger liquid) fixed point with gapless excitations in
$\phi_{+\rho}$. For commensurate density $n=p/q$, $H_{\rm{eff}}$
will also contain $H_{2q}$ if the latter is a relevant operator with
respect to $H_{LL}$. In that case $H_{2q}$ will lock $\phi_{+\rho}$
in one of the minima of its potential, thereby opening up a gap in
$\phi_{+\rho}$, causing true long-range order. Assuming that
$\langle A_{2q}(x)\rangle \neq 0$ for the field-locking pattern in
the gapped sectors (if $\langle A_{2q}(x)\rangle$ should happen to
vanish, true long-range order is ruled out), the only decay of the
two-point function of $H_{2q}\equiv \int dx\;{\cal H}_{2q}(x)$ comes
from the $\exp{(\pm 2iq\phi_{+\rho})}$ factors, giving the
asymptotic decay $\langle {\cal H}_{2q}(x){\cal H}_{2q}(0)\rangle
\sim |x|^{-2d_{2q}}$, where
\begin{equation}
d_{2q}=q^2 K_{+\rho} \label{scaldim}
\end{equation}
is the scaling dimension of $H_{2q}$. If $d_{2q}<2$, $H_{2q}$ is
perturbatively relevant. Thus the critical value of $K_{+\rho}$
below which $H_{2q}$ is expected to open up a gap and cause true
long-range order is from this simple argument estimated to be
\begin{equation}
K_{+\rho}^c \equiv 2/q^2. \label{Kc}
\end{equation}
The dopings considered in Sec. \ref{dmrg} correspond to very large
values of $q$, and consequently very small values of $K_{+\rho}^c$
(see Table \ref{scaldimtab}). For weak repulsive interactions
$K_{+\rho}$ is expected to be slightly less than 1 ($K_{+\rho}=1$ in
the noninteracting model). Therefore $H_{2q}$ is strongly irrelevant
for weak interactions in the doped ladder, and consequently true
long-range order is not possible in that case. For larger
interactions a definite statement can no longer be made, as
$K_{+\rho}$ is not known as a function of $U$, $V_{\perp}$,
$J_{\perp}$ etc. On the other hand, true long-range order would
certainly be very surprising. Although one generally expects
$K_{+\rho}$ to decrease with increasing interaction strength, the
values of $K_{+\rho}^c$ for the nonzero dopings in Table
\ref{scaldimtab} are, to the best of our knowledge, orders of
magnitude smaller than the smallest known values of $K_{+\rho}$ in
the $t$-$J$ and Hubbard ladders,\cite{smallK} which furthermore are
obtained only for very strong interactions, much stronger than the
ones used in Sec. \ref{dmrg}. Correspondingly, only for low
commensurabilities (i.e., small values of $q$) has it been found
that $K_{+\rho}$ can become as small as $K_{+\rho}^c$ for sufficiently
strong interactions in theoretical ladder models, with
true long-range order as a result.
\begin{table}
\begin{center}
\begin{tabular}{|c|c|c|l|}\hline
$\delta$ (\%) & $p$ & $q$ & $K_{+\rho}^c$ \\\hline
0 & 1 & 1 & 2 \\
2 & 49 & 50 & 0.0008 \\
2.86 & 373 & 384 & 0.0000136 \\
4, 8, 12 & 24, 23, 22 & 25 & 0.0032 \\
\hline
\end{tabular}
\caption{The parameters $p$, $q$, and $K_{+\rho}^c=2/q^2$ evaluated
for the rational dopings $\delta=1-p/q$ considered in Sec.
\ref{dmrg}. The values of these parameters at half-filling are shown
for comparison.} \label{scaldimtab}
\end{center}
\end{table}

We end this subsection with some remarks on how the discussion
presented here relates to previous work. It is already known in the
literature that RG/bosonization arguments predict that having true
long-range order in the two-leg ladder requires relevant Umklapp
interactions and that Umklapp interactions are usually irrelevant
away from half-filling. These conclusions, and arguments leading to
them, have been invoked either explicitly or implicitly in many
previous studies of interacting electrons on a two-leg ladder (see,
e.g., Refs.
\onlinecite{sch96,balfis96,origia97,linbalfis97,linbalfis98,gognertsv98,sch99,scawhiaff01,fjamar02,whiaffsca02,
tsufur02,oricit03,wuliufra03,gia03}). The discussion given here
based on symmetry considerations, which is a generalization to the
two-leg ladder of previous discussions for a single
chain,\cite{aff90,sch94,gia97} allowed for a more systematic and
general derivation of these results. In particular, the derivation
of Eq. (\ref{scaldim}) enabled us to give a semi-quantitative
discussion of the question of \textit{how} irrelevant (or relevant)
the most relevant allowed Umklapp operator is for an arbitrary
rational filling.

\subsection{Quasi long-range order scenario}
\label{qlroscen}

The weak-coupling analysis in the previous subsection suggests that
Umklapp interactions are unlikely to be relevant for the dopings
considered in Sec. \ref{dmrg}, and that as a consequence the
effective theory at sufficiently low energies and long wavelengths
is given by the Luttinger liquid Hamiltonian in Eq. (\ref{HLutt}),
with gapless excitations in $\phi_{+\rho}$. This will be referred to
as the quasi long-range order scenario, as it implies that for
sufficiently large distances correlation functions of exponentials
of $\phi_{+\rho}$ decay as power laws with $K_{+\rho}$-dependent
exponents (the same features hold for $\theta_{+\rho}$). In this
subsection we summarize some consequences of this scenario, both for
infinite and finite ladders. From a numerical comparison with the
Luttinger-liquid predictions for finite ladders, we find that the
DMRG results are not consistent with the quasi long-range order
scenario. This analysis complements the discussions in Secs.
\ref{dmrg} and \ref{tlroscen} which concluded that the DMRG results
are consistent with a true long-range order scenario.

\begin{table*}[h,t]
\begin{center}
\begin{tabular}{|c||ll|ll|ll|}\hline
 Correlation & \multicolumn{6}{|c|}{Phase}\\\cline{2-7}
 function & \multicolumn{2}{c|}{SF/DDW} & \multicolumn{2}{c|}{CDW} & \multicolumn{2}{c|}{Doped D-Mott} \\\hline\hline
 $(2k_F,\pi)$-SF/DDW           & $K_{+\rho}/2$\cite{wuliufra03} & (const) & exp & (exp) & exp & (exp) \\
 $(2k_F,\pi)$-CDW & exp & (exp) & $K_{+\rho}/2$\cite{wuliufra03} & (const) & exp & (exp) \\
 DSC & exp & (exp) & exp & (exp) & $1/(2K_{+\rho})$\cite{sch96,sch99,whiaffsca02,wuliufra03} & (exp) \\
 $(4k_F,0)$-CDW   & $2K_{+\rho}$ & (const) & $2K_{+\rho}$ & (const) &
 $2K_{+\rho}$\cite{sch96,sch99,whiaffsca02} & (const) \\\hline
\end{tabular}
\caption{Bosonization predictions for the asymptotic decay of
two-point correlation functions in the two-leg ladder. In each cell,
the first entry is the decay when $\phi_{+\rho}$ is gapless (quasi
long-range order scenario); the second entry (in parenthesis) is the
decay when $\phi_{+\rho}$ is gapped (true long-range order
scenario). Power-law decay is indicated by the decay exponent,
`const' means no decay, and `exp' means exponential decay.}
\label{decaytable}
\end{center}
\end{table*}

We start by summarizing the Luttinger-liquid predictions for the
correlation functions for the charge density, orbital current and
DSC pairing operators in the doped SF/DDW, CDW and D-Mott phases in
an infinitely long ladder. The nature of the decay of these
correlation functions is easily obtained from the bosonized
expressions in Sec. \ref{physobsdoped}, the field-locking patterns
for the $\pm\sigma$ and $-\rho$ sectors (Table \ref{phasetable}),
and the correlation functions of exponentials of $\phi_{+\rho}$ and
$\theta_{+\rho}$. The results are summarized in Table
\ref{decaytable}. A few remarks are in order. The $4k_F$ density
correlations show the same qualitative behavior across the entire
phase diagram. In the SF/DDW and CDW phases, the SF/DDW and
$2k_F$-CDW correlations, respectively, dominate over the $4k_F$-CDW
correlations. In the doped D-Mott phase, the DSC correlations are
dominant for $K_{+\rho}>1/2$ while the $4k_F$-CDW correlations
dominate for $K_{+\rho}<1/2$.\cite{sch96,sch99,whiaffsca02}
Schulz\cite{sch99} has argued that $K_{+\rho}\to 1$ as $\delta\to
0^+$ in the doped D-Mott phase. Therefore one expects DSC
correlations to be dominant in the doped D-Mott phase sufficiently
close to half-filling. (For easy comparison the nature of the
asymptotic decay when $\phi_{+\rho}$ is gapped has been indicated in
parenthesis in Table \ref{decaytable}. The cases we have commented
on here are those for which the true and quasi long-range order
scenarios predict qualitatively different decays.)

The expressions for the power-law exponents in Table
\ref{decaytable} spur the question of whether the doped ladder might
be described by a quasi long-range order scenario with a very small
$K_{+\rho}$. Unfortunately, the strong finite-size effects in the
DMRG results severely complicate a comparison with the
Luttinger-liquid predictions for an infinite ladder. It is much
preferred to compare the DMRG results with analytical predictions
for a ladder of \textit{finite} size and with \textit{open} boundary
conditions. White {\em et al.}\cite{whiaffsca02} have recently
discussed such a ladder that is in a Luttinger-liquid phase of the
type described above, i.e., in which the low-energy effective theory
is given by the gapless Luttinger-liquid Hamiltonian $H_{\text{LL}}$
in the $+\rho$ sector and the three other sectors are gapped. It was
shown that in such a system the boundaries will act as impurities
and induce generalized Friedel oscillations in the charge density
which decay away from the boundaries. These Friedel oscillations
have fundamental wavevector $4k_F$ and thus period $1/\delta$. White
{\em et al.} furthermore showed that the density amplitude in the
middle of the ladder should scale with the ladder length $L$ as
$L^{-K_{+\rho}}$. To investigate whether the $4k_F$ charge density
oscillations found in Sec. \ref{dmrg} might be interpreted as
generalized Friedel oscillations in such a Luttinger liquid, one can
plot the mid-ladder density amplitude as a function of the ladder
length in a log-log plot and try to fit the results to a straight
line; from its slope one can extract a value for
$K_{+\rho}$.\cite{whiaffsca02}

We have attempted this kind of fit of the DMRG results for two
different cases, both at 4 percent doping: In the SF/DDW phase
($t=V_{\perp}=1$, $U=0.25$ and $J_{\perp}=1.5$), and in the doped
D-Mott phase (same parameters except $J_{\perp}=1.8$). In both cases
the mid-ladder density amplitude is found not to decay with ladder
length for sufficiently large ladders, so that the fit gives
$K_{+\rho}=0$. This result does not support an interpretation of the
density oscillations as generalized Friedel oscillations in a
Luttinger liquid; instead it is consistent with a true long-range
order scenario. Even if one were to take into account corrections to
$H_{\text{LL}}$ due to irrelevant operators (such as the
basic-Umklapp interaction discussed more below), one would still
expect that the mid-ladder density amplitude should decrease
monotonically towards zero with increasing ladder length, instead of
approaching a nonzero constant. Thus if the DMRG results reflect the
true behavior of the system the quasi long-range order scenario
appears to be ruled out.

\subsection{Neglecting quantum fluctuations: A classical description of
the doped ladder}
\label{effmodel}

The fact that the basic-Umklapp interaction is relevant in the
half-filled ladder enabled true long-range order in that case. The
analysis in Sec. \ref{condtrue} suggests that in the doped ladder
the basic-Umklapp interaction is irrelevant: although it can cause
significant ordering tendencies on length scales $\lesssim
O(1/\delta)$, it can be neglected on length scales $\gg
1/\delta$.\cite{objection} The Luttinger-liquid behavior discussed
in Sec. \ref{qlroscen}, with quasi long-range ordered correlation
functions, would thus be expected to manifest itself in this
asymptotic regime.\cite{wuliufra03} One would think that this regime
should be accessible numerically since the dopings considered in
Sec. \ref{dmrg} correspond to values of $1/\delta$ that are
considerably smaller than the largest system sizes studied there. In
spite of these expectations, we have seen that the DMRG results are
not consistent with quasi long-range order.

In this section we will show that a good qualitative description of
the DMRG results can in fact be obtained from a purely classical
treatment of a low-energy effective Hamiltonian describing the
effects of doping the half-filled phase, a phase for which we have
seen that the basic-Umklapp interaction plays a crucial role. The
classical model is obtained by neglecting, by hand, the quantum
fluctuations in $\phi_{+\rho}$ that if kept would have caused
$\phi_{+\rho}$ to disorder on length scales $\gg 1/\delta$.

\subsubsection{Classical model for the SF/DDW phase}

To derive the classical model for the SF/DDW phase, we take as our
starting point the low-energy effective Hamiltonian for this phase
at half-filling, which will be referred to as $\mathscr{H}_0$. Its
form is assumed to be well approximated by the form for weak
interactions, Eqs. (\ref{H0boson})-(\ref{HI2}), with values of the
effective couplings appropriate for the system being in the SF/DDW
phase for intermediate-strength interactions. In order to
investigate the effects of doping, we add to $\mathscr{H}_0$ a
chemical potential term $\Delta \mathscr{H}=-\mu {\cal N}$, where
\begin{equation}
{\cal N}\equiv \sum_{P\lambda s}{\cal N}_{P\lambda s}=\frac{2}{\pi}\int dx\; \partial_x \phi_{+\rho}
\end{equation}
measures the number of electrons in the system with respect to the
half-filled case (note that here we work with a fixed chemical
potential $\mu$ instead of at a fixed doping $\delta$ as we have
done up until now both in the DMRG and bosonization analyses). The
fields $\phi_{+\sigma}$, $\theta_{-\rho}$, and $\theta_{-\sigma}$
are taken to be locked to values consistent with being in the SF/DDW
phase, and the resulting gaps will be assumed to be large enough
that excitations in these fields may be safely neglected when
discussing the low-energy physics.

To obtain an effective, classical Hamiltonian from $\mathscr{H}_0$
we proceed as follows: (i) Neglect all terms that do not contain
$+\rho$ fields, as these can be regarded as constants for our
purposes. (ii) In the basic-Umklapp term (\ref{HI2}), replace $\cos
2\phi_{+\sigma}$, $\cos 2\theta_{-\rho}$, $\cos 2\phi_{-\sigma}$,
and $\cos 2\theta_{-\sigma}$ by their expectation values (which will
generally depend on $\mu$). This gives an effective coupling
constant for $\cos 2\phi_{+\rho}$. (iii) Neglect the term containing
($\partial_x \theta_{+\rho})^2$ which gives rise to quantum
fluctuations in $\phi_{+\rho}$. This makes $\phi_{+\rho}$ a purely
classical field which allows us to neglect the distinction between
$\phi_{+\rho}$ and $\langle \phi_{+\rho}\rangle$ in the following.
Then the effective classical Hamiltonian, denoted by
$\mathscr{H}_{+\rho}$, can be written
\begin{equation}
\mathscr{H}_{+\rho}\equiv \int_0^L dx\;\left[A\left(\frac{d\phi_{+\rho}}{dx}\right)^2-B\cos 2\phi_{+\rho}\right],
\label{Ham}
\end{equation}
where $A$ and $B$ are effective positive $\mu$-dependent coupling
constants. Since we are here considering the theory on a finite
system size $L$, it should be noted that boundary effects have not
been taken into account in this effective Hamiltonian.

We want to consider a system with a given average doping
$\delta={\cal N}_h/2L$, where ${\cal N}_h=-{\cal N}$ is the number
of doped holes. Thus $\Delta \mathscr{H}$ is simply a constant
$-2|\mu|\delta L$ which can be neglected, and $\phi_{+\rho}$ has to
satisfy the boundary condition
\begin{equation}
\phi_{+\rho}(L)-\phi_{+\rho}(0)= -\pi\delta L.
\label{phibc}
\end{equation}
Since the net spin in the $z$ direction is measured by the operator
\begin{equation}
\sum_{P\lambda s}s\,{\cal N}_{P\lambda s}=\frac{2}{\pi}\int_0^L dx\;\partial_x \phi_{+\sigma},
\end{equation}
the fact that $\langle \partial_x \phi_{+\sigma}\rangle = 0$ in the
low-energy subspace implies that the number of up- and down-spin
electrons are equal. Consequently, up- and down-spin electrons are
removed in pairs by the chemical potential, so that ${\cal N}_h$ is
an even integer, and therefore the product $\delta L$ in Eq.
(\ref{phibc}) is an integer. Note that this condition is satisfied
for all figures discussed in Sec. \ref{dmrg} except Fig.
\ref{fig:holes11} where the presence of an unpaired hole is seen to
result in disturbances in the density and current patterns.

The problem we have arrived at consists in finding the function
$\phi_{+\rho}(x)$ which minimizes $\mathscr{H}_{+\rho}$, subject to
the boundary condition (\ref{phibc}). That is, the solution
$\phi_{+\rho}(x)$ is the classical ground state configuration of the
sine-Gordon model in the presence of a finite density of
solitons;\cite{arilut02} equivalently, we can think of it as a
Frenkel-Kontorova-like problem.\cite{chalub95} There is a
competition between $\mathscr{H}_{+\rho}$ which wants to lock
$\phi_{+\rho}$ in one of the minima of the cosine potential, and the
boundary condition which forces $\phi_{+\rho}$ to have a nonzero
slope.

Once $\phi_{+\rho}(x)$ has been found, the currents and densities
are easily obtained from Eqs. (\ref{js}) and
(\ref{rungcurarb})-(\ref{densityarb}). It is important to note that
since the doping $\delta$ is determined by the term $-\mu {\cal N}$
in the effective Hamiltonian, the $\delta=0$ form of Eqs.
(\ref{rungcurarb})-(\ref{densityarb}) must be used. Consequently,
the staggered rung current can be written
\begin{equation}
j_s(x) = j_0 \cos \phi_{+\rho},
\label{jdoped}
\end{equation}
where $j_0 \approx j_{\pi} \langle \cos \phi_{+\sigma} \cos
\theta_{-\rho} \cos \theta_{-\sigma}\rangle \neq 0$. The value of
the amplitude $j_0$ will be determined by fitting to the DMRG
results. Furthermore, the electron density on the top and bottom
site at rung $x$ are equal and given by
\begin{equation}
n(x)=1 + \frac{1}{\pi}\frac{d\phi_{+\rho}}{dx}.
\label{densop}
\end{equation}

\subsubsection{Solution of the model}
\label{solmodel}

To solve this problem analytically, we interpret
$\mathscr{H}_{+\rho}$ as the classical {\em action} for a particle
with mass $2A$, located at position $\phi$ at time $x$, with
potential energy $B\cos 2\phi$ (here and in the rest of this
paragraph we omit the subscript $+\rho$ on $\phi_{+\rho}$). It will
be convenient to take the potential energy to be $V(\phi) \equiv
B(\cos 2\phi-1)=-2B\sin^2 \phi\leq 0$ corresponding to a
redefinition of its zero. The sum of kinetic and potential energies
is conserved,
\begin{equation}
A \left(\frac{d\phi}{dx}\right)^2+V(\phi)=E,
\label{encons}
\end{equation}
where $E$ is the total energy of the particle. According to Eq.
(\ref{phibc}), in the time interval $L$ the particle moves a
distance $\pi \delta L$, which (since $L>1/\delta$) is bigger than
$\pi$, the distance between the potential minima. Thus the particle
is not trapped inside one of the wells of the potential $V(\phi)$,
but is unbounded. Hence  $E>0$, so that the velocity $d\phi/dx$
always has the same sign, which must be negative since
$\phi(L)<\phi(0)$. Eq. (\ref{encons}) can then be integrated to give
\begin{equation}
x-x_0=-\sqrt{a}\;F(\phi \;|-2b),
\end{equation}
where $a\equiv A/E$, $b\equiv B/E$, $F(\phi|m)$ is the incomplete
elliptic integral of the first kind (see Appendix \ref{elliptic}),
and $x_0$ is an integration constant defined by $\phi(x_0)\equiv 0$.
This gives
\begin{equation}
\phi(x) = \mbox{am}\bigg(-\frac{x-x_0}{\sqrt{a}}\bigg|-2b\bigg),
\label{phi}
\end{equation}
where $\mbox{am}(u|m)$ is the Jacobian amplitude function. Using
Eqs. (\ref{phibc}), (\ref{phi}), and (\ref{amprop}), and the
assumption that $L$ is an integer multiple of $1/\delta$, the
parameters $a$ and $b$ can be related as
\begin{equation}
\frac{1}{\sqrt{a}}=2\delta\,K(-2b),
\label{eqE}
\end{equation}
where $K(m)$ is the complete elliptic integral of the first kind.

The staggered rung current $j_s(x)$ and electron density $n(x)$ can
then be expressed in terms of the Jacobian elliptic functions cn and
dn, respectively:
\begin{eqnarray}
\hspace{-1cm} j_s(x) &=& j_0 \;\mbox{cn}\bigg(\frac{x-x_0}{\sqrt{a}}\bigg|-2b\bigg), \label{rungcur} \\
\hspace{-1cm} n(x) &=& 1 - \frac{1}{\pi\sqrt{a}} \;\mbox{dn}\bigg(\frac{x-x_0}{\sqrt{a}}\bigg|-2b\bigg).
\label{rungden}
\end{eqnarray}
Using Eq. (\ref{eqE}) to eliminate $a$, we see that $n(x)$ can be
expressed in terms of the unknown parameters $b$ and $x_0$, while
for $j_s(x)$ an additional parameter $j_0$ is also needed. The
parameter $x_0$ plays the role of a phase variable which
distinguishes between different solutions related by translation
along the $x$ axis, which all have the same energy since we have
neglected boundary effects.

Using Eqs. (\ref{eqE})-(\ref{rungden}) and the properties of the
elliptic functions, the following results are easily established:
$n(x)$ and $j_s(x)$ oscillate with wavelengths $1/\delta$ and
$2/\delta$, respectively, and $n(x)$ is minimal where $j_s(x)$
changes sign, i.e. at the anti-phase domain walls in the current
pattern. We emphasize that these properties of the analytical
solution are independent of the actual values of the fitting
parameters $b$, $j_0$ and $x_0$, and are in excellent agreement with
the numerical DMRG results.

\subsubsection{Fits to DMRG results}

\begin{figure}
\centerline{\includegraphics[scale=0.75]{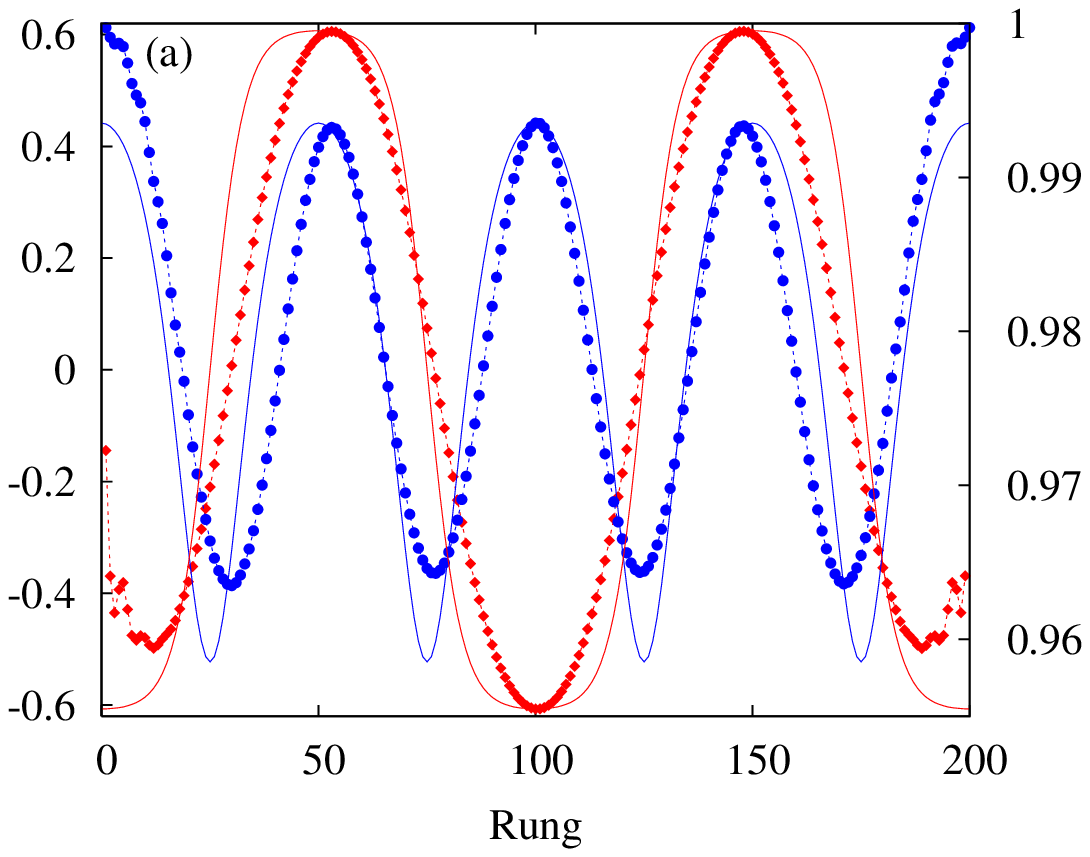}}
\centerline{\includegraphics[scale=0.75]{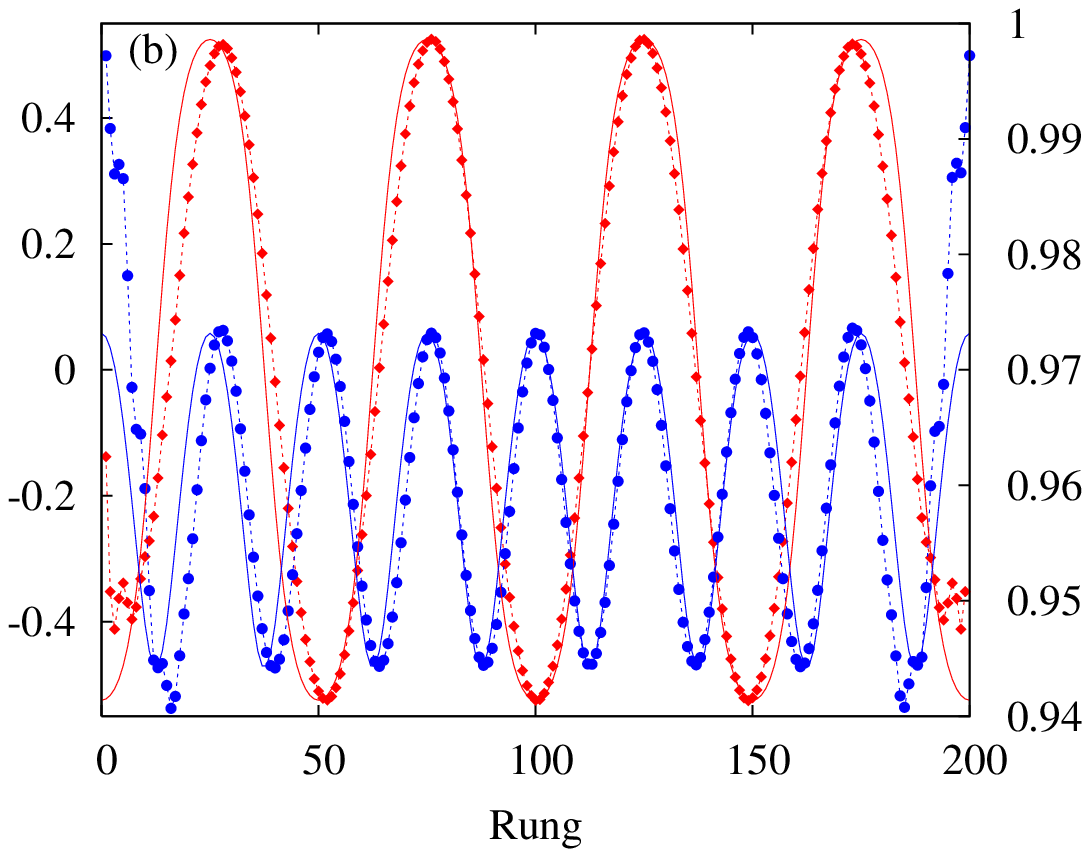}}
\hspace{-1.7cm}\centerline{\includegraphics[scale=0.75]{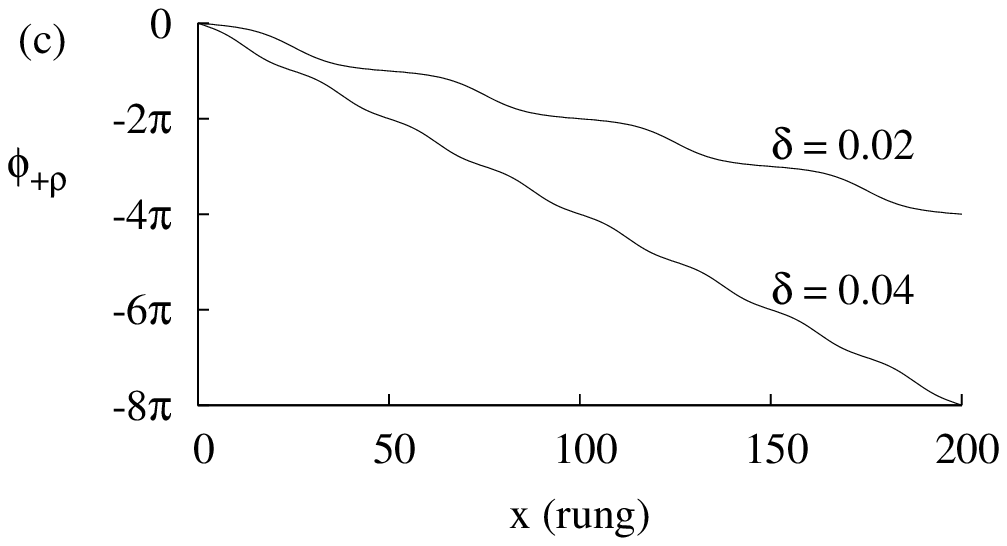}}
\caption{(Color online) (a) Staggered rung current $j_s(x)$ (red
curves, left y-axis) and electron density per site $n(x)$ (blue
curves, right y-axis) as a function of rung index $x$ in the SF/DDW
phase of a 200-rung ladder with $\delta=0.02$. Shown are DMRG
results (diamonds and circles for currents and densities,
respectively, connected by dashed lines to guide the eye) and
analytical fits (full lines). The model parameters for the DMRG
calculation are $t=V_{\perp}=1$, $U=0.25$, $J_{\perp}=1.5$ and
$J_{\parallel}=V_{\parallel}=0$. The values of the fitting
parameters used are $b\approx 20.0$, $|j_0|\approx 0.61$, and
$x_0=0$. (b) Plots for the same model parameters as above, except
that the doping $\delta=0.04$. The values of the fitting parameters
are $b\approx 1.66$, $|j_0|\approx 0.52$, and $x_0=0$. (c) The
function $\phi_{+\rho}(x)$ corresponding to the fits of $j_s(x)$ and
$n(x)$.} \label{qLRO-fit}
\end{figure}

In Fig. \ref{qLRO-fit} we show fits of the analytical expressions
for $j_s(x)$ and $n(x)$ to DMRG results for a 200-rung ladder for
$\delta=0.02$ (upper panel) and $\delta=0.04$ (middle panel). The
corresponding solutions of $\phi_{+\rho}(x)$ are also shown (lower
panel). For both dopings, the model parameters, and therefore the
DMRG results, are identical to those in Fig. \ref{fig:trueLROfit}.
The fit is rather good for $\delta=0.04$, the main deviations
occurring near the boundaries due to the neglect of boundary effects
in the analytical solution. However, since by construction both the
numerical and analytical curves for $n(x)$ integrate to half the
total number of electrons, deviations between the density curves
near the boundaries inevitably imply some deviations also away from
the boundaries. These deviations come mainly in the form of a
horizontal shift between the density curves which decreases towards
the center of the ladder. There is an identical shift between the
current curves so that the zeros of the current are always at the
minima of the density. It is as if the boundaries exert a ``push''
on the density and current oscillations, and this effect cannot be
captured by the analytical solution.

Similar features are observed in the fit for the lower doping
$\delta=0.02$, but now there are also more pronounced differences
between the shapes of the analytical and numerical curves. In the
analytical fit the current oscillations have a more rectangular
shape and the density minima are deeper. It is natural to attribute
these differences primarily to the omission of quantum fluctuations
in the effective model. The step-like nature of the solitons in the
analytical solution for $\phi_{+\rho}(x)$ becomes more pronounced
with decreasing doping $\delta$ (see lower panel in figure), as the
ratio of the soliton separation $1/\delta$ to the soliton width
becomes larger. We expect that quantum fluctuations will be more
efficient in smoothing the solitons the more step-like these
solitons are. This picture is consistent with the reduced quality of
the fit as the doping is reduced from $0.04$ to
$0.02$.\cite{comment-fit-2} Comparison with the DMRG results in Fig.
\ref{qLRO-fit} thus seems to suggest that while the solitons in
$\phi_{+\rho}(x)$ would not be much affected by the inclusion of
quantum fluctuations in the effective model for $\delta=0.04$, they
would be somewhat smoothed for $\delta=0.02$, leading to less
rectangular current oscillations and not so deep minima in the
electron density.

\subsubsection{The CDW and doped D-Mott phase}

The results for the charge density oscillations in the CDW and doped
D-Mott phase that follow from such a classical approach are also in
good qualitative agreement with the DMRG results. In the CDW phase,
the density on leg $\ell$ of rung $x$ takes the form (obtained from
Eq. (\ref{densityarb}) with $\delta=0$)
\begin{equation}
n_{t,b}(x)=1+\frac{1}{\pi}\frac{d\phi_{+\rho}}{dx} \pm
\tilde{n}_0(-1)^{x}\cos\phi_{+\rho}
\end{equation}
where $\tilde{n}_0\approx n_{2k_F}\langle \cos
\phi_{+\sigma}\sin\theta_{-\rho} \cos \theta_{-\sigma}\rangle\neq
0$. In the doped D-Mott phase the density also has this form except
that $\tilde{n}_0=0$. In both phases the solution for $\phi_{+\rho}$
is given by Eq. (\ref{phi}), as in the SF/DDW phase. Thus $(4k_F,0)$
density oscillations with period $1/\delta$, coming from the
$d\phi_{+\rho}/dx$ term, are predicted to be present in all phases
(for the CDW phase, the DMRG results show that the $(2k_F,\pi)$
density oscillations dominate over the $(4k_F,0)$ oscillations in
size, however). Furthermore, because DMRG suggests that
$\phi_{+\rho}$ essentially behaves like a classical field over the
length scales considered here, the conjugate field $\theta_{+\rho}$
is strongly fluctuating, so that DSC correlations decay
exponentially over these length scales even in the doped D-Mott
phase.

\section{Summary and discussion}
\label{conclusion}

In this paper we have studied a generalized Hubbard model on a
two-leg ladder, focusing on a parameter region that shows SF/DDW
order. The approximate location of this region in parameter space
was found from a calculation of the phase diagram of the half-filled
model for weak interactions, which was done by using Abelian
bosonization and semiclassical considerations to analyze the
low-energy effective theory resulting from a one-loop RG flow.

The RG/bosonization calculations in this and previous
works\cite{tsufur02,schetal03,wuliufra03} suggest that in order to
get SF/DDW order in the generalized Hubbard ladder it is necessary
(although not sufficient) to have the on-site repulsion $U$ be less
than the nearest neighbor repulsion $V_{\perp}$. One class of
materials that may be promising for achieving a relatively small $U$
is organic molecular crystals, due to the large size of the highest
occupied molecular orbital; bringing two of these molecules close
together might then conceivably give $U\lesssim V$.\cite{ben}
Another way of reducing the effective $U$ is through an on-site
(Holstein) electron-phonon interaction.\cite{bervallin95}

To study (rational) hole dopings $\delta$ away from half-filling,
finite-system DMRG calculations were done on ladders with up to $2
\times 200$ sites for intermediate-strength interactions. It was
found that the SF/DDW phase persists up to quite high dopings (of
order 10-20\%) and that upon doping, currents remain large, with no
evidence of decay. The rung currents oscillate with wavevector
$(2k_F,\pi)$, corresponding to oscillations with wavelength
$2/\delta$ in the {\em staggered} rung current. The currents coexist
with $(4k_F,0)$ charge density oscillations, with two doped holes
per wavelength $1/\delta$, which also are not found to decay.

The factor of two ratio between the periodicities of the staggered
rung current and the rung charge density, and the related fact that
the maxima of the hole density are located at the anti-phase domain
walls of the rung current, imply that the SF/DDW phase is an example
of a phase with ``topological doping."\cite{kiveme96} In this
respect, the rung current in the SF/DDW phase is the analogue of the
spin density in the so-called ``stripe phases" in doped
antiferromagnets, which have coexistence of spin and charge density
order.\cite{caremekivorg02}

Charge density oscillations which do not seem to decay are also
found in the neighboring CDW and doped D-Mott phases, in which
currents decay exponentially. Both these phases have $(4k_F,0)$
charge density oscillations; the CDW phase in addition has
$(2k_F,\pi)$ charge density oscillations. DSC correlations decay
exponentially in all phases (except possibly deep inside the doped
D-Mott phase where the numerical results for the decay are more
difficult to interpret).

We have shown that most of these DMRG results for the doped ladder
can be qualitatively understood from weak-coupling RG/bosonization
arguments. However, there is one apparent exception: The DMRG
results are consistent with true long-range order in the currents in
the SF/DDW phase and true long-range order charge density order in
all phases. As far as we know, such a true long-range order scenario
is not forbidden by any exact theorems that are directly applicable
to the lattice Hamiltonian studied here; in particular, the
Mermin-Wagner theorem is respected. On the other hand, the
RG/bosonization arguments imply that such a scenario would require
the symmetric charge field $\phi_{+\rho}$ to become locked and that
this can only happen if multiple-Umklapp interactions are relevant
in the RG sense. However, a calculation of the scaling dimension of
these interactions suggests that they are irrelevant for the dopings
considered, and thus the true long-range order scenario does not
appear to be supported by the RG/bosonization arguments. These
arguments further suggest a crossover from non-decaying
current/charge density correlations to power-law decay at a length
scale of order $1/\delta$, due to the basic-Umklapp interaction
being unimportant on much longer length scales.\cite{wuliufra03} For
the same reasons, in the doped D-Mott phase a crossover from
exponential to power-law decay in the DSC correlations would be
expected at the same characteristic length scale.

In light of this discrepancy between the predictions of the DMRG and
RG/bosonization method for the long-range order question, a detailed
scrutiny of both methods might be necessary, to see if one could
possibly identify an omission or shortcoming of the conventional
analysis that, if corrected, would resolve the discrepancy. While
any investigations along these lines are beyond the scope of the
present work, we will in the next paragraphs comment on the issues
that might appear most salient.

For the DMRG method, a standard concern is whether boundary and/or
finite-size effects could cause the system to appear to have true
long-range order even if there in reality is none. However, the
boundaries are in fact found to suppress SF/DDW order, not enhance
it. Furthermore, the result of a finite-size test for Luttinger
liquid behavior (quasi long-range order), that involved the
evolution of the mid-ladder density with ladder length, was
negative. Thus there seems to be no reason to expect the DMRG
results to change qualitatively for larger systems than we have
studied here. The DMRG calculations are also ``internally
consistent" in the sense that the DSC correlations in the doped
D-Mott phase are found to be consistent (or at least not
inconsistent) with exponential decay, in agreement with the
bosonization result when $\phi_{+\rho}$ is locked.

As the true long-range order scenario implies that $\phi_{+\rho}$ is
gapped, it would have been desirable to use DMRG to calculate the
charge gap directly, as function of ladder length, and see if it
extrapolates to a nonzero value in the thermodynamic
limit.\cite{whiaffsca02} However, this calculation is very difficult
and we have unfortunately not been able to do it.

It has been known for a long time \cite{ostrom95,sch04} that DMRG
works with ansatz states of the form of matrix product
states.\cite{fannacwer92,kluschzit93} Within this class of ansatz
states it finds, with some qualification of no importance to the
present argument, the variationally optimal state as approximation
to the true ground state. One might think that this methodological
bias might sometimes algorithmically favor a state that is not a
good approximation to the true ground state (in the present case,
suggesting long-range instead of quasi long-range order). To our
knowledge there is however no known case where converged DMRG
calculations have seriously misrepresented the true quantum state
where it is known from other methods. Moreover, matrix product
states would typically bias more in favor of shorter-ranged
correlations. Yet, this scenario cannot be totally excluded.

One concern about the RG/bosonization method is to what extent its
predictions can be trusted when the interactions are not weak. Wu et
al. have recently argued that the quasi long-range order predictions
would be expected to be valid also for the interaction strengths
used in the DMRG calculations presented here although the parameters
in the low-energy effective theory might be strongly
renormalized.\cite{wuliufra03} A concrete example of the validity
and usefulness of RG/bosonization arguments for analyzing a ladder
with strong interactions was given in Ref. \onlinecite{whiaffsca02}.
In particular, for $3/8$ filling in the $t$-$J$ ladder with
$J/t\approx 0.25$, DMRG calculations showed evidence for a charge
density wave with true long-range order; the values for $K_{+\rho}$
extracted from the DMRG results for $J/t\gtrsim 0.25$ were
consistent with the critical value $K_{+\rho}^c=0.125$ predicted by
RG/bosonization for the onset of true long-range
order.\cite{whiaffsca02} But this example does of course not imply
correctness of the weak-coupling RG/bosonization predictions in the
model considered by us.

Another, much more speculative concern is whether some of the RG
arguments used in this paper might break down in a more fundamental
way. A scenario for the breakdown of the perturbative RG at certain
quantum critical points has recently been discussed in Ref.
\onlinecite{belkirrol04}. The breakdown occurs because the expansion
coefficients of terms that appear in the RG equations at
\textit{two-loop} order depend singularly on a dangerously
irrelevant variable, which causes them to diverge. Physically, the
effect is related to the presence of another diverging time scale in
addition to the critical one. For coupled one-dimensional chains,
nonlinear corrections to the dispersion (i.e., band curvature
terms), which like the Umklapp interactions reflect the underlying
lattice structure of the problem, do seem to behave in a way that
might be characterized as dangerously irrelevant with respect to
some transport properties (Coulomb drag).\cite{pusetal03,gru03}
Whether as a result of this something akin to the breakdown
described in Ref. \onlinecite{belkirrol04} might occur in the model
studied here is an intriguing, but at this stage highly speculative,
question.

Although resolving the long-range order question is an important
theoretical issue, as a practical matter it may make little
difference whether there is true long-range order or only algebraic
order that decays very slowly. Physical realizations of systems such
as we consider in this paper will either involve weakly coupled
ladders, or a full two-dimensional lattice. If isolated ladders can
only show quasi-long-range ordered current and charge density
correlations, even extremely weak coupling between such ladders
could suffice to stabilize true long-range order. We note that
recent results of inelastic neutron scattering measurements on
Sr$_{14}$Cu$_{24}$O$_{41}$ have been interpreted as a possible
signature of orbital currents.\cite{bouetal04} Clearly it would be
very interesting if further experiments on this material were to
corroborate this interpretation.

\begin{acknowledgments}
We are grateful to I. Affleck, F. F. Assaad, M. A. Cazalilla, S.
Chakravarty, F. H. L. Essler, E. Fradkin, A. Furusaki, T. Giamarchi,
I. A. Gruzberg,
A. F. Ho, S. A. Kivelson, P. A. Lee, R. H. McKenzie, A. Paramekanti,
B. J. Powell, T. Senthil and M. Troyer for helpful and stimulating
discussions. J. O. F. gratefully acknowledges support for this work
from funds from the David Saxon chair at UCLA. For the time during
which this work was being completed, he thanks the Australian
Research Council for financial support, and the Rudolf Peierls
Centre for Theoretical Physics at Oxford University for its
hospitality. J. B. M. was supported in part by the US National
Science Foundation, Grant Nos. DMR-0213818 and PHY99-0794. He also
thanks the Aspen Center for Physics for its hospitality during a
stay there in the summer of 2003, and MIT and the Kavli Institute
for Theoretical Physics for extended stays in 2004.
\end{acknowledgments}

\appendix

\section{Klein factor conventions}
\label{kleinconv}

Here we explain the conventions used for the Majorana Klein
factors\cite{schcunpie98,sen04} in the bosonized versions of the
Hamiltonian and the various order parameters considered in this
paper.

The nonquadratic part of the Hamiltonian density, ${\cal
H}_{I}^{(1b)}+{\cal H}_I^{(2)}$ in Eqs. (\ref{HI1b})-(\ref{HI2}),
contains the Hermitian operator
$\Gamma\equiv\eta_{1\up}\eta_{1\down}\eta_{2\up}\eta_{2\down}$.
Furthermore, the bosonized order parameters for the SF/DDW and CDW
phases, Eqs. (\ref{SF/DDW-OP})-(\ref{CDW-OP}), contain the two
anti-Hermitian operators $h_{s}\equiv \eta_{2s}\eta_{1s}$,
$s=\uparrow,\downarrow$, while the bosonized order parameters for
DSC and SSC correlations, Eqs. (\ref{DSC-OP})-(\ref{SSC-OP}),
contain the two anti-Hermitian operators
$h'_{\lambda}\equiv\eta_{\lambda\uparrow}\eta_{\lambda\downarrow}$,
$\lambda=1,2$. From $\Gamma^2=-h_{s}^2=-h'^{2}_{\lambda}=1$, it
follows that $\Gamma$ has eigenvalues $\pm 1$, while $\{h_{s}\}$ and
$\{h'_{\lambda}\}$ have eigenvalues $\pm i$.

Since
$[\Gamma,h_{s}]=[\Gamma,h'_{\lambda}]=[h_{s},h_{s'}]=[h'_{\lambda},h'_{\lambda'}]=0$,
it is possible to simultaneously diagonalize the order parameter of
a given phase and the Hamiltonian in the Klein factor space. (Note
that although $[h_{s},h'_{\lambda}]\neq 0$, this causes no problem
for such a simultaneous diagonalization, because each order
parameter contains either $\{h_{s}\}$ or $\{h'_{\lambda}\}$, not
both.)

A consistent choice of eigenvalues can then be deduced from the
relations $\Gamma=-h_{\uparrow}h_{\downarrow}=h'_1 h'_2$. In this
paper we choose the eigenvalues $\Gamma=1$, $h_{s}=i$, and
$h'_{\lambda}=i(-1)^{\lambda}$.

\section{Renormalization group equations and initial conditions for continuum
couplings at half-filling}
\label{apprg}

The one-loop RG equations can be derived using the operator product
expansion (OPE).\cite{car95,balfis96,linbalfis97,sen04} At
half-filling they read
\begin{eqnarray}
\dot{g}_{1\rho} &=& g_{t\rho}^2+\frac{3}{16}g_{t\sigma}^2 -
g_{tu1}^2-g_{tu2}^2-g_{tu1}g_{tu2}, \nonumber \\
\dot{g}_{x\rho} &=& -g_{t\rho}^2-\frac{3}{16}g_{t\sigma}^2 - g_{xu}^2
-g_{tu1}^2-g_{tu2}^2-g_{tu1}g_{tu2}, \nonumber \\
\dot{g}_{1\sigma} &=& -g_{1\sigma}^2 - \frac{1}{2}g_{t\sigma}^2 +
2g_{t\rho}g_{t\sigma}-4g_{tu1}^2 - 4g_{tu1}g_{tu2}, \nonumber \\
\dot{g}_{x\sigma} &=& -g_{x\sigma}^2-\frac{1}{2}g_{t\sigma}^2 -
2g_{t\rho}g_{t\sigma} - 4g_{tu2}^2 - 4g_{tu1}g_{tu2}, \nonumber \\
\dot{g}_{t\rho} &=& 2g_{t\rho}(g_{1\rho}-g_{x\rho})+\frac{3}{8}
g_{t\sigma}(g_{1\sigma}-g_{x\sigma}) \nonumber \\ &-& g_{xu}(g_{tu1}-g_{tu2}), \nonumber \\
\dot{g}_{t\sigma} &=& 2 g_{t\rho}(g_{1\sigma}-g_{x\sigma})+
g_{t\sigma}(2g_{1\rho}-2g_{x\rho}-g_{1\sigma}-g_{x\sigma}) \nonumber \\
 &+& 4 g_{xu}(g_{tu1}+g_{tu2}), \nonumber \\
\dot{g}_{xu} &=& -4 g_{x\rho}g_{xu}-g_{tu1}\left(2g_{t\rho}-\frac{3}{2}
g_{t\sigma}\right) \nonumber \\ &+& g_{tu2}\left(2g_{t\rho}+\frac{3}{2}g_{t\sigma}\right), \nonumber \\
\dot{g}_{tu1} &=& -g_{tu1}\left(2g_{1\rho}+2g_{x\rho}+\frac{3}{2}g_{1\sigma}
-\frac{1}{2}g_{x\sigma}\right) - g_{1\sigma}g_{tu2} \nonumber \\ &+& g_{xu}\left(-2 g_{t\rho}
+\frac{1}{2}g_{t\sigma}\right), \nonumber \\
\dot{g}_{tu2} &=& -g_{tu2}\left(2g_{1\rho}+2g_{x\rho}-\frac{1}{2}g_{1\sigma}
+\frac{3}{2}g_{x\sigma}\right)-g_{x\sigma}g_{tu1} \nonumber \\ &+& g_{xu}\left(2g_{t\rho}
+\frac{1}{2}g_{t\sigma}\right).
\label{RGeqs}
\end{eqnarray}
Here $\dot{g}_i \equiv 2\pi v_F dg_i/dl$, where $l$ is related to
the running cutoff scale as $\alpha(l)=\alpha e^l$. These RG
equations were first derived in Ref. \onlinecite{balfis96}. The
initial values for the couplings are found to be
\begin{eqnarray}
g_{1\rho} &=& -\frac{1}{4}U-\frac{1}{4}V_{\perp}-\frac{5}{4}V_{\parallel}+\frac{3}{16}J_{\perp}
-\frac{3}{16}J_{\parallel}, \nonumber \\
g_{1\sigma} &=& U+V_{\perp}-V_{\parallel}- \frac{3}{4}J_{\perp}-\frac{3}{4}J_{\parallel}, \nonumber \\
g_{x\rho} &=& -\frac{1}{4}U-\frac{3}{4}V_{\perp}-\frac{3}{2}V_{\parallel}
-\frac{3}{16}J_{\perp}-\frac{3}{8}J_{\parallel}, \nonumber \\
g_{x\sigma} &=& U-V_{\perp}-2V_{\parallel}-
\frac{1}{4}J_{\perp}-\frac{1}{2}J_{\parallel}, \nonumber \\
g_{t\rho} &=& -\frac{1}{4}U+\frac{1}{4}V_{\perp}-V_{\parallel}-\frac{3}{16}J_{\perp}-\frac{3}{8}J_{\parallel},
\nonumber \\
g_{t\sigma} &=& U-V_{\perp}-2V_{\parallel}+\frac{3}{4}
J_{\perp}, \nonumber \\
g_{xu} &=& -\frac{1}{2}U+\frac{1}{2}V_{\perp}+V_{\parallel}-\frac{3}{8}
J_{\perp}-\frac{3}{4}J_{\parallel}, \nonumber \\
g_{tu1} &=& -\frac{1}{2}U-\frac{1}{2}V_{\perp}+\frac{1}{2}V_{\parallel}-\frac{1}{8}J_{\perp}-\frac{5}{8}J_{\parallel},
\nonumber \\
g_{tu2} &=& \frac{1}{2}U-\frac{1}{2}V_{\perp}-V_{\parallel}-
\frac{1}{8}J_{\perp} + \frac{1}{2}J_{\parallel}.
\label{initcond}
\end{eqnarray}
Eqs. (\ref{RGeqs}) and (\ref{initcond}) hold for $t_{\perp}=t$, i.e.
$k_{F2}=2k_{F1}=2\pi/3$. As a partial check of the internal
consistency of these equations, it is instructive to generalize them
to arbitrary values of $t_{\perp}/t$ (though $<2$ so that both bands
are occupied) and by also allowing for density-density and spin
exchange interactions along the plaquette diagonals with strength
$V'$ and $J'$, respectively.\cite{fjaunpub} It can then be shown
that the generalized equations have the correct symmetries in
special limiting cases: U(4) symmetry when $t_{\perp}=0$,
$U=V_{\perp}$, $V_{\parallel}=V'$, $J_{\parallel}=J_{\perp}=J'=0$;
U(2)$\times$U(2) symmetry when
$t_{\perp}=V_{\perp}=V'=J_{\perp}=J'=0$ (i.e. independent legs), and
SO(5) symmetry when\cite{scazhahan98,linbalfis98}
$J_{\perp}=4(U+V_{\perp})$, $V_{\parallel}=V'=J_{\parallel}=J'=0$.
(Note that when $t_{\perp}=0$ three new continuum couplings are
allowed in addition to the nine present for a generic $t_{\perp}$.)

\section{Elliptic integrals of the first kind and associated Jacobian elliptic functions}
\label{elliptic}

In this Appendix we summarize notation\cite{ellnot} for and some
basic properties of the elliptic integrals of the first kind and the
associated Jacobian elliptic functions that are encountered in the
analytical solution of the Frenkel-Kontorova-like problem in Sec.
\ref{solmodel}.

The definitions of the incomplete and complete elliptic integrals of the first kind are, respectively,
\begin{eqnarray}
F(\phi|m) &\equiv & \int_0^{\phi}\frac{d\alpha}{\sqrt{1-m \sin^2 \alpha}}, \label{incell} \\
K(m) &\equiv & F(\pi/2|m). \label{compell}
\end{eqnarray}
We will only be concerned with negative values of $m$ here. The
inverse function of $u\equiv F(\phi|m)$ exists and is defined as
\begin{equation}
\phi\equiv\mbox{am}(u|m).
\label{am}
\end{equation}
This function is known as the amplitude for the Jacobian elliptic functions.
It is odd in $u$, and satisfies
\begin{equation}
\mbox{am}(u+2K(m)|m)=\mbox{am}(u|m)+\pi.
\label{amprop}
\end{equation}
The Jacobian elliptic functions encountered in our problem are
\begin{eqnarray}
\mbox{cn}(u|m) &\equiv& \cos\phi, \label{cn} \\
\mbox{dn}(u|m) &\equiv& \frac{d\phi}{du} = \sqrt{1-m \sin^2 \phi}. \label{dn}
\end{eqnarray}
These two functions are even and periodic in $u$ with period $4K(m)$ for cn and $2K(m)$ for dn.


\begin{thebibliography}{}
\bibitem{DWterm} The name ``d-density wave" originates from the particle-hole nature of the condensate and the
$d_{x^2-y^2}$ symmetry of its singlet order parameter; see Ref. \onlinecite{nay00}.
\bibitem{phasenames} The names D-Mott and S-Mott were introduced in Ref. \onlinecite{linbalfis98}. Note that in
Ref. \onlinecite{schetal03} the D-Mott and S-Mott phase were referred to as the rung singlet and site singlet
phase, respectively.
\bibitem{DMottdoped} The ``Mott'' in the name ``doped D-Mott phase'' is not meant to imply that the phase that
results when doping the D-Mott phase is necessarily insulating.
Whether or not the various rationally doped phases are insulating is
one of the central questions discussed in this paper.
\bibitem{checkumklapp} The form of the (most relevant) Umklapp
operator has already been considered in the literature for some
special commensurate fillings: $n=1/2$
(quarter-filling),\cite{whiaffsca02,oricit03} $n=3/4$ and
1/4,\cite{whiaffsca02} and of course $n=1$ (half-filling). Our
result for $H_{2q}$ agrees in all cases.
\bibitem{ph-breaking} Due to the absence of particle-hole symmetry there is an additional term proportional to
$v_{F1}-v_{F2}$ in the continuum model away from half-filling. Following previous treatments, we assume that
this term is irrelevant, at least for sufficiently weak dopings.
\bibitem{smallK} The smallest values of $K_{+\rho}$ we are aware of in the literature occur in connection with a
possible charge density wave phase with true long-range order in the
$t$-$J$ ladder for $n=0.75$ and $J/t\lesssim 0.25$ (i.e.
corresponding to very strong interactions), for which some evidence
was recently presented in Ref. \onlinecite{whiaffsca02}. In this
case, it is expected that $K_{+\rho}\to 0.125$ as $J/t$ is reduced
towards its  critical value $(J/t)_c$ for $n=0.75$, and that
$K_{+\rho}\to 0.0625$ as $n\to 0.75$ for
$J/t<(J/t)_c$.\cite{whiaffsca02}
\bibitem{objection} One possible objection to this conclusion is that the analysis in Sec. \ref{condtrue} was
based on weak-coupling arguments, and that the classification of the basic-Umklapp interaction as irrelevant
might therefore not be valid for intermediate and strong interactions. However, as pointed out in Ref.
\onlinecite{whiaffsca02}, although the quantities $k_{F1}$ and $k_{F2}$ may be renormalized by interactions
(and may not even be well defined in the interacting model), their sum is protected by the one-dimensional
version of Luttinger's theorem,\cite{yamoshaff97,gaghaaric98} so that the relation $4k_F=2\pi(1-\delta)$ is
expected to be exact. This suggests that the conclusions regarding the basic-Umklapp interaction should hold
more generally. This should also include the order-of-magnitude estimate $1/\delta$ for the length scale
signifying the crossover from insulating behavior at shorter length
scales to metallic behavior at longer length scales.
\bibitem{comment-fit-2} We note that by choosing a much smaller value of the fitting parameter $b$ in Fig.
\ref{qLRO-fit}, the fit of the current could have been much
improved, but the density oscillations in the analytical curve would
have been smaller than in the numerical curve, which would not be
consistent with our picture of the role of quantum fluctuations.
\end{thebibliography}
\end{document}